\documentclass[journal=ancac3,manuscript=article,layout=twocolumn]{achemso}
\usepackage{inputenc}
\usepackage{xcolor}
\usepackage{graphicx}
\usepackage{amsmath}
\usepackage{amssymb}
\usepackage{array}
\usepackage{booktabs}
\usepackage{listings}
\usepackage{lmodern}
\usepackage{mathpazo}
\usepackage{microtype}
\usepackage{caption}
\usepackage{float}
\usepackage{geometry}
\usepackage{natbib}
\usepackage{setspace}
\usepackage{xkeyval}

\newcommand{\ket}[1]{\big|#1\big>}

\newcommand{\bk}{\mathbf{k}}
\newcommand{\bd}{\mathbf{d}}

\newcommand{\bq}{\mathbf{q}}

\def\dz2{d$_{\text{z}^2}$}

\def\dx2y2{d$_{\text{x}^2\text{y}^2}$}
\def\G0W0{G$_0$W$_0$} 
\def\scGW0{scGW$_0$}

\def\mos{MoS$_2$}
\def\mose{MoSe$_2$}
\def\ws{WS$_2$}
\def\wse{WSe$_2$}

\title{Optical nonlinearities in the excited carrier density of atomically thin transition metal dichalcogenides}

\author{D. Erben}
\affiliation{Institute for Theoretical Physics, University of Bremen, P.O. Box 330 440, 28334 Bremen, Germany}
\email{derben@itp.uni-bremen.de}

\author{A. Steinhoff}
\affiliation{Institute for Theoretical Physics, University of Bremen, P.O. Box 330 440, 28334 Bremen, Germany}

\author{M. Lorke}
\affiliation{Institute for Theoretical Physics, University of Bremen, P.O. Box 330 440, 28334 Bremen, Germany}

\author{F. Jahnke}
\affiliation{Institute for Theoretical Physics, University of Bremen, P.O. Box 330 440, 28334 Bremen, Germany}

\begin{document}
\begin{abstract}
In atomically thin semiconductors based on transition metal dichalcogenides, photoexcitation can be used to generate high densities of electron-hole pairs. Due to optical nonlinearities, which originate from Pauli blocking and many-body effects of the excited carriers, the generated carrier density will deviate from a linear increase in pump fluence. In this paper, we use a theoretical approach that combines results from ab-initio electronic-state calculations with a many-body treatment of optical excitation to describe nonlinear absorption properties and the resulting excited carrier dynamics. We determine the validity range of a linear approximation for the excited carrier density vs. pump power and identify the role and magnitude of optical nonlinearities at elevated excitation carrier densities for MoS$_2$,  MoSe$_2$,  WS$_2$, and WSe$_2$ considering various excitation conditions. We find that for above-band-gap photoexcitation, the use of a linear absorption coefficient of the unexcited system can strongly underestimate the achievable carrier density for a wide range of pump fluences due to many-body renormalizations of the two-particle density-of-states.
\end{abstract}
\section{Introduction}
Monolayers of transition metal dichalcogenide (TMDC) semiconductors exhibit strong Coulomb interaction of their charge carriers that occupy a rich valley structure in reciprocal space.  As a new degree of freedom these valleys are selectively optically addressable \cite{xu_spin_2014,ugeda_giant_2014}, giving rise to a variety of Coulomb-bound, bright and dark excitons, trions and  biexcitons.\cite{qiu_optical_2013,chernikov_exciton_2014,ye_probing_2014, tang_long_2019, plechinger_trion_2016,florian_dielectric_2018, hao_neutral_2017, steinhoff_biexciton_2018}
Furthermore, giant band-gap renormalization \cite{steinhoff_influence_2014,meckbach_giant_2018}, efficient carrier scattering\cite{kaasbjerg_hot-electron_2014, steinhoff_nonequilibrium_2016, danovich_fast_2017}, and the Mott transition of excitons at elevated carriers densities \cite{steinhoff_exciton_2017} have been discussed.
For a better understanding of the underlying physics and the prospects of TMDC semiconductors as active materials in future optoelectronic devices like light-emitting diodes \cite{pospischil_solar-energy_2014, baugher_optoelectronic_2014, ross_electrically_2014, withers_light-emitting_2015},  solar  cells \cite{pospischil_solar-energy_2014, baugher_optoelectronic_2014}  or  lasers \cite{wu_monolayer_2015, ye_monolayer_2015, salehzadeh_optically_2015, li_room-temperature_2017,lohof_prospects_2019},  experimental techniques such as photoluminescence and pump-probe spectroscopy are used. \cite{he_tightly_2014,steinhoff_efficient_2015,klots_probing_2014,chernikov_population_2015,wang_optical_2019} These experiments involve photoexcitation of electron-hole pairs, frequently above the quasi-particle band gap, and benefit from reliable estimates of the present excited carrier density.
The simplest approach to the excitation density is based on the frequency-dependent linear absorption coefficient. \cite{sie_observation_2017,bataller_dense_2019,li_slow_2019} Normalizing the pump fluence to the laser spot size and taking the absorption coefficient at the excitation energy together with the photon energy yields a reasonable approximation for low intensities. However, as soon as the exciting laser pulse is intense enough, effects such as phase space filling are expected to reduce the absorption in a nonlinear way.  Furthermore, excited carriers will induce screening of the Coulomb interaction, thereby reducing excitonic effects and modifying interband Coulomb enhancement in the interacting density of states entering the optical absorption. Therefore, it is desirable to quantify the involved nonlinearities microscopically and calculate the resulting excitation densities for typical experimental situations.
In this paper we provide a numerical analysis of the photo-excited charge carrier dynamics as well as optical nonlinearities in absorption and excited carrier density. Our approach combines material-realistic electronic state and interaction matrix element calculations with semiconductor Bloch equations including many-body effects. 

\section{Photoexcited charge carrier densities in TMDCs}\label{sec:results}
\begin{figure*}[h!t]
\centering
\includegraphics[trim =0cm 8cm 0cm 8cm, clip, width=\textwidth]{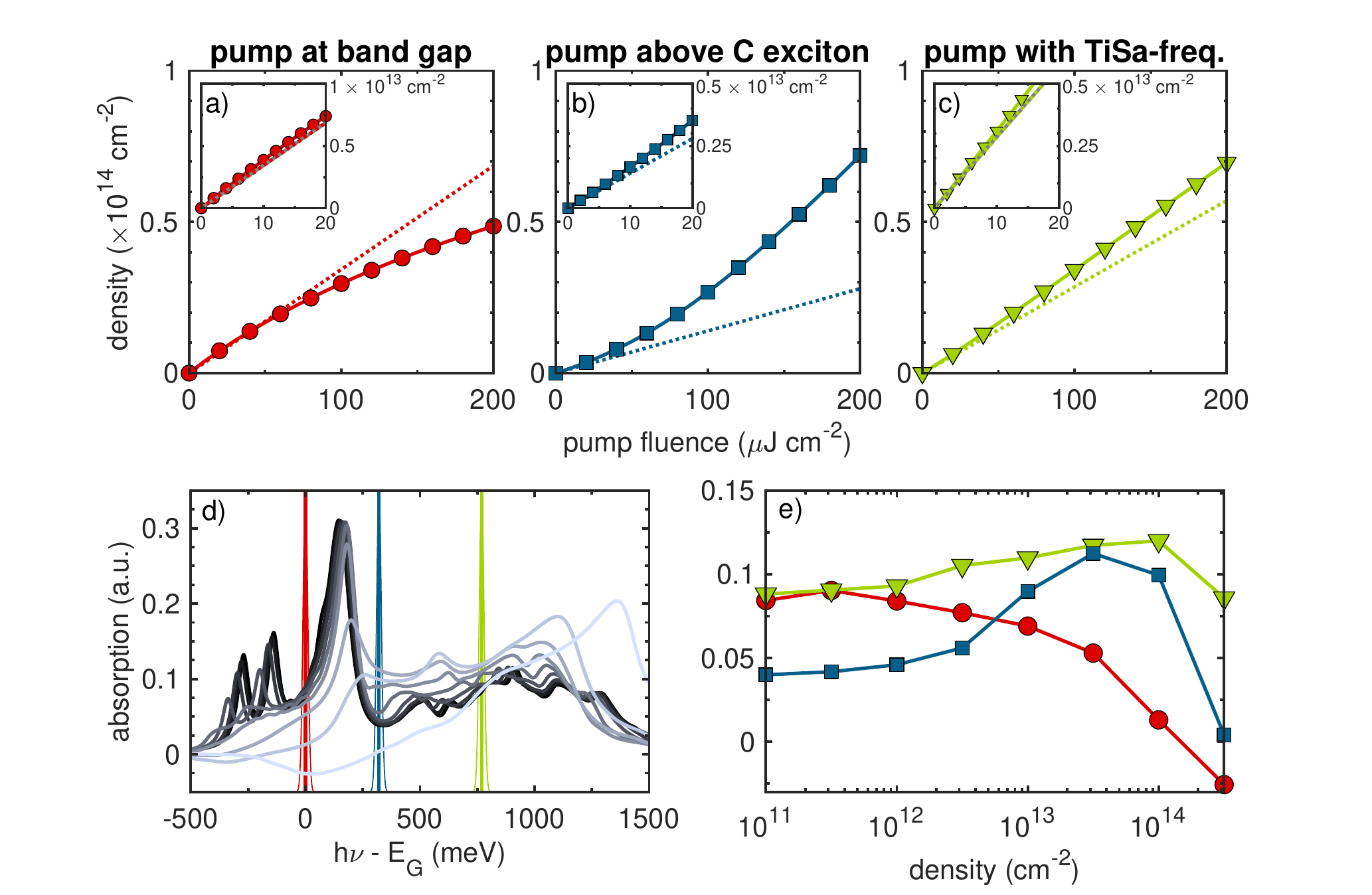}
\caption{Optically excited charge carrier density vs. pump fluence for excitation of MoS$_2$ with a 150 fs laser pulse tuned to the band-gap energy (a), above the C exciton (b), and well above band gap corresponding to the Ti-sapphire laser emission wavelength (c). Solid lines and symbols represent calculated carrier densities including optical nonlinearities. Dashed lines correspond to carrier densities obtained from the calculated linear absorption coefficients of the unexcited system at the respective energies. Absorption spectra of MoS$_2$ for a given excited carrier density in thermal equilibrium at 300K are shown in (d) together with the energetic position and spectral width of the pump pulses corresponding to (a)-(c). The carrier density increases from zero (black line) to $3 \times 10^{14}$ cm$^{-2}$ (gray line) corresponding to the data points in (e). The nonlinear behaviour of the extracted absorption coefficients vs. carrier density at the three pump energies is depicted in (e). 
}
\label{fig:fig2}
\end{figure*}
\begin{figure*}[h!t]
\centering
\includegraphics[trim =0cm 8.5cm 0cm 8.5cm, clip, width=\textwidth]{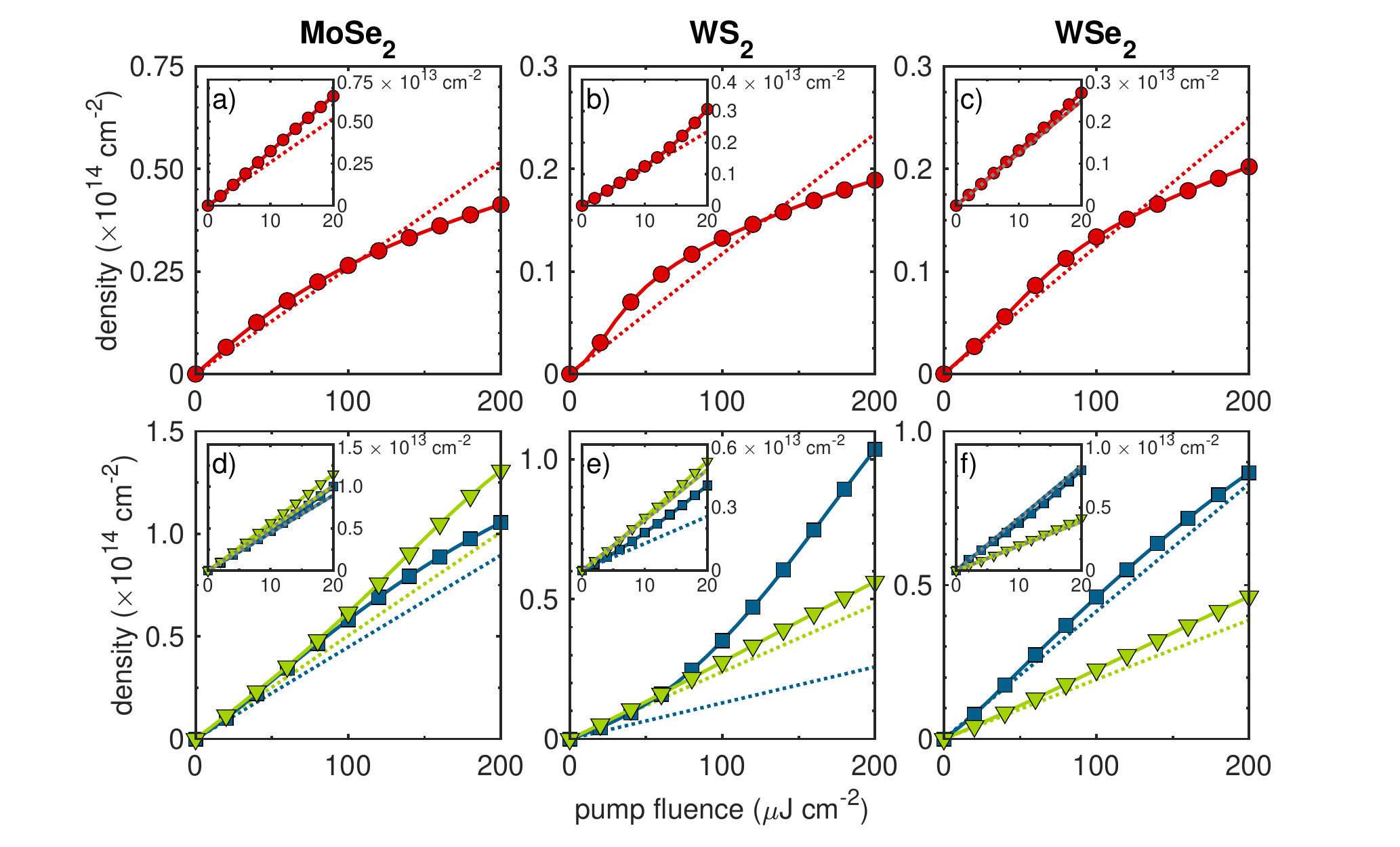}
\caption{Optically excited charge carrier density vs. pump fluence for excitation of MoSe$_2$ (left), WS$_2$ (center), and WSe$_2$ (right) with a 150 fs laser pulse tuned to the band-gap energy (red), above the C exciton (blue), and well above band gap corresponding to the TiSa emission wavelength (green). Solid lines and symbols represent calculated carrier densities including optical nonlinearities. Dashed lines correspond to carrier densities obtained from the calculated linear absorption coefficients of the unexcited system at the respective energies.
}
\label{fig:fig3}
\end{figure*}

A microscopic theory for the photoexcited charge carrier density in TMDC semiconductors requires addressing several subtopics and suitable interfacing of the methods used in these parts. For electronic state calculations, we start from density functional theory (DFT) with many-body perturbation theory corrections in GW approximation and perform a basis set reduction to the relevant part of the band structure by constructing a six-band lattice Hamiltonian.\cite{steinhoff_influence_2014} Four conduction and two valence bands are used to describe the dominant contributions to the photoexcitation process and the subsequent excited carrier dynamics. Light-matter interaction and the resulting excited carrier dynamics are treated using the semiconductor Bloch equations (SBE)\cite{steinhoff_influence_2014,haug_quantum_2004} which include excitonic effects, Pauli-blocking due to excited carriers, energy renormalizations (shifts of the band structure) due to the Coulomb interaction of excited carriers, as well as carrier scattering (equilibration and thermalization) processes. The SBE can be used for calculating photoexcited carrier densities and optical absorption spectra under the influence of the above discussed effects. The theory includes a consistent calculation of Coulomb and dipole interaction matrix elements, the description of screening effects (background lattice contributions including encapsulation layers and excited carrier contributions), as well as dephasing and light propagation effects in the system. Details are provided in the Methods section.\\
The density of photoexcited charge carriers at a specific fluence is obtained from a time-domain numerical solution of the SBE into steady state without the presence of recombination processes. Assuming that the excited-carrier dynamics is faster than typical recombination processes, our results reflect the achievable photoexcitation densities in a time window between pulsed excitation and recombination.\\
Our findings for the dependence of excited-carrier density on detuning and pump fluence are provided in Figs.~\ref{fig:fig2} and \ref{fig:fig3}. 
For better interpretation of the results, we also show calculated linear absorption spectra for different constant densities of photoexcited carriers in thermal equilibrium at room temperature. Only for these calculations of the spectra, the SBE are solved for fixed quasi-equilibrium occupation probabilities $F_{\bk}^{e/h}$ and a weak electric probe field is used to extract the linear optical susceptibility of the TMDC including many-body effects. The susceptibility is translated into absorption using the transfer matrix method for a thin layer in an hBN environment.\cite{jahnke_linear_1997} We note, that for optical pumping considered in Fig.~\ref{fig:fig2} (a) - (c) and Fig.~\ref{fig:fig3}, photoexcited carrier density and nonequilibirum population functions $f^{e/h}_\bk(t)$ vary continuously during the pulse, which is fully considered in the dynamical calculations.
\\
Results for \mos{} are collected in Fig. \ref{fig:fig2}. For pumping at the single-particle band gap, a nonlinear relation between density and fluence is found, which can be directly related to the Pauli blocking effect at the band edge. Nevertheless, this effect is partially compensated by quasi-particle energy renormalizations that spectrally shift the band gap away from the pump pulse to lower energies, thereby opening up phase space at the pump energy. This leads to a nearly linear increase of the excited carrier density for fluences up to $20$ $\mu$J cm$^{-2}$ and corresponding carrier densities up to $10^{13}$ cm$^{-2}$. As shown in the inset of Fig.~\ref{fig:fig2} (a), band-structure renormalizations even overcompensate phase space filling in this fluence range, so that a slighty superlinear increase of the density is achieved.
For pumping above the band gap, the linear regime extends only to $5$ $\mu$J cm$^{-2}$ and excited-carrier densities are typically larger than for pumping at the band gap. This is a result of the reduced Pauli blocking as the carrier scattering rapidly distributes excited carriers away from the pump laser resonance to lower energies.
In comparison to the carrier densities obtained from extrapolating the zero-density linear absorption (dotted lines), larger values are obtained. We attribute this to increasing absorption at the pump laser resonance with increasing excited carrier density, as can be seen in Fig.~\ref{fig:fig2} (d) and (e). Energy renormalization shifts regions of larger interband density-of-states from higher energies $h\nu$ onto the pump laser resonance. Furthermore, the increasing dephasing also contributes to the effect.
\\
A similar behaviour for pumping at the single-particle band gap is found for other TMDC materials, as shown in Fig.~\ref{fig:fig3}. While the superlinear increase of carrier density due to overcompensation of the phase-space-filling reduction by energy renormalization is quantitatively comparable in the Mo-based materials, the initial overshooting extends to larger pump fluences for the W-based materials. This effect originates from stronger spin-orbit interaction: the splitting between A and B excitons is increased so that the B exciton is located above the single-particle gap at zero density. Under photoexcitation, band-gap renormalization, screening, and Pauli blocking cause a red shift and bleaching of A and B excitons. The remnant of the B exciton is shifted into resonance with the pump laser for excitation at the single-particle band gap, thereby providing an additional boost of the two-particle density-of-states, see the density-dependent absorption spectra of Fig.~S1 in the Supporting Information.
For above band-gap pumping, the dependence of photoexcited carrier density on fluence is superlinear already at fluences of several $\mu$J cm$^{-2}$ for the sulfides, while it is closer to or even below the linear estimate for the selenides. This can be traced back to the single-particle band gap of the selenides being smaller by about $0.3$ eV, see Tab.~S1. As a consequence, pumping above the C exciton or at the Ti-sapphire laser frequency drives band-to-band transitions that are correspondingly further above the gap. These spectral regions profit less from a renormalization of the two-particle density-of-states than in the sulfides, see Fig.~S1 and Fig.~S2 in the Supporting Information.
\\
It should be noted, that the calculated excited carrier densities in Figs.~\ref{fig:fig2} and \ref{fig:fig3} follow from a self-consistent solution of the SBE. This solution includes coupled polarization and population dynamics and, as such, does not use the absorption spectrum of the system. For the used parameters and excitation conditions, the system is close to the adiabatic regime, where the carrier density evolution follows approximately the pulse area. During different stages of the pump pulse, the population increase is approximately determined by the absorption of the system for the momentary carrier distribution.
Correspondingly, the absorption spectrum of the system in Fig.~\ref{fig:fig2} (d) and Fig.~S1 can be used to interprete the results. In the realized incoherent regime with ultra-fast equilibration of carriers, absorption changes during the pump pulse reflect the excited-carrier nonlinearities.
\begin{figure*}[h!t]
\centering
\includegraphics[trim =0cm 11cm 0cm 11.5cm,clip,width=\textwidth]{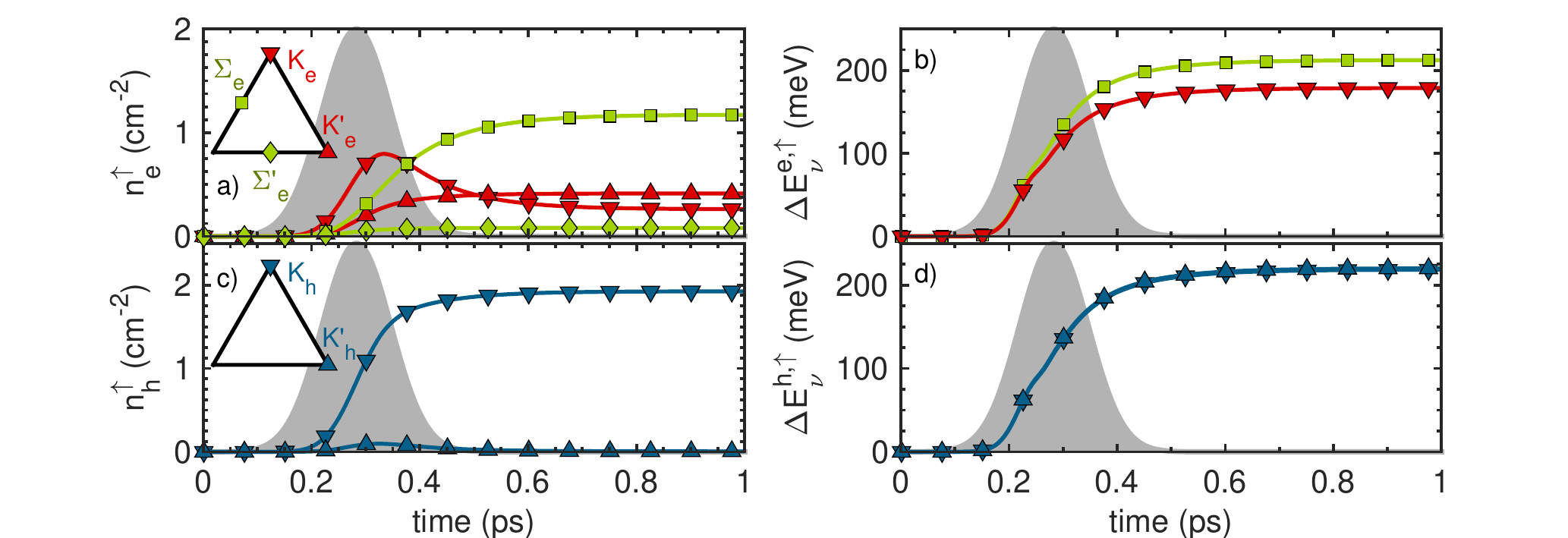}
\caption{Time evolution of the optically excited carrier density in \mos{} for spin-up electrons (a) and holes (c) resolved for the different valleys in the Brillouin zone when pumping at the band gap. The corresponding band-structure renormalizations $\Delta E_\nu^\lambda = |\varepsilon_\nu^\lambda(0) - \varepsilon_\nu^\lambda(t)|$ for these valleys are provided in (b) and (d). Curves for K$'_e$ and $\Sigma'_e$ have been omitted for better visibility as they show the same overall behaviour as K$_e$ and $\Sigma_e$ except that $\Sigma'_e$ is renormalized weaker than K$'_e$. The excitation pulse is represented by the gray shaded area.
}
\label{fig:fig4}
\end{figure*}
We also note, that radiative coupling of the induced interband polarization $\mathbf{P}_{\textrm{TMDC}}(t)$ within the TMDC layer reduces the effective electric field $\mathbf{E}(t)$ according to Eq.~\ref{eq:effective_E}, see Methods section. Thus, less coherence is converted into photoexcited carrier density. The net effect is approximately a 10\% excitation-density reduction.
\\
Additional insight into the build-up of the excited-carrier densities in the different valleys for pumping at the single-particle band gap is provided in Fig.~\ref{fig:fig4}. Panels (a) and (c) show the rise of carrier density in the most important band-structure valleys (K, K', $\Sigma$, $\Sigma'$). The large densities at $\Sigma/\Sigma'$ reflect a drain of electrons from the K-valley, where they are excited by the pump pulse. This is due to the valley energy renormalizations induced by the excited carriers, see Fig.~\ref{fig:fig4} (b) and (d). A transition to an indirect band gap takes place, as the conductions-band $\Sigma$-valley experiences a stronger energy lowering than the K-valley and eventually the $\Sigma$-valley energy is below that of the $K$-valley\cite{erben_excitation-induced_2018}. Since the band-structure renormalizations almost directly follow the pump pulse and the relaxation of carriers, carrier drain takes place faster than typical radiative recombination times, which are on the order of $1-10$ ps\cite{fang_control_2019} depending on substrate and temperature\cite{palummo_exciton_2015}. Loss of carriers into the $\Sigma$-valley has been associated with a sublinear increase of photoluminescence with pump fluence\cite{steinhoff_efficient_2015}.
\section{Conclusion}\label{sec:conclusion}
This paper provides a critical analysis, to which extent linear absorption coefficients can be used to determine the excited carrier density in TMDC semiconductors. A combination of material-realistic electronic-state calculations with a theory for optical properties under the influence of excited-carrier many-body effects has been used to model optical pulse excitation and the subsequent carrier dynamics. Results are different for pumping at or above the quasi-particle band gap. In the former case, a linear regime is found for surprisingly large excited carrier densities up to $1 \times 10^{13}$ cm$^{-2}$ for \mose{} and up to $2 \times 10^{12}$ cm$^{-2}$ for \mos{}, \ws{}, and \wse{} due to a compensation of Pauli-blocking nonlinearities in the absorption and band-structure renormalizations. For above-gap excitation, excited carrier densities can strongly exceed the linear extrapolation values since absorption reduction due to Pauli blocking remains weak while stronger absorption regions at higher energies can be shifted onto the pump resonance due to giant band-gap shrinkage.
\section{Methods}
\subsection{Microscopic theory of photoexcited TMDC monolayers}\label{sec:theory}
A microscopic description of TMDC semiconductors photoexcited with intense short laser pulses needs to adress various aspects:
(i) The material specific properties of the band structure as well as dipole and Coulomb interaction matrix elements. The latter include dielectric screening effects due to the TMDC lattice as well as possible encapsulation layers and exhibit a specific momentum dependence owing to the atomically thin material \cite{chernikov_exciton_2014}.
(ii) The dynamics of the light-matter interaction following a short-pulse excitation and subsequent carrier relaxation. The effects are reflected in the excited carrier population dynamics.
(iii) Many-body interaction effects of the excited carriers with a variety of contributions leading to energy renormalizations of the electronic single-particle states, excitonic effects, carrier scattering processes, as well as dephasing influencing optical properties and providing dissipation in the coherent excitation dynamics.
(iv) Light-propagation effects, where the applied electromagnetic laser field induces a coherent material polarization in the atomically thin TMDC layer which acts back on the electromagnetic field thus leading to an effective field within the material.\cite{jahnke_linear_1997}
While the comprehensive solution of the full problem is a challenging task, the goal of this paper is to use suitable approximation schemes for the individual problems and construct interfaces between treatments of the subproblems.
\\
Our treatment is based on well-established ab-initio electronic-state calculations of TMDC monolayers using the DFT-GW scheme. The excited carrier dynamics will be described with semiconductor Bloch equations (SBE). \cite{haug_quantum_2004,steinhoff_influence_2014} To interface both approaches, we use a Wannier function lattice Hamiltonian with a reduced basis set involving the relevant bands for direct coupling to the optical field. This interface provides material-realistic dipole and Coulomb interaction matrix elements. In the coherent excitation dynamics, excitonic effects are included by solving the SBE with these matrix elements entering the electron-hole interaction terms.
\\
For the treatment of many-body effects we focus on above-bandgap excitation, as used in many experiments. A typical scenario involves two-pulse excitation, where the intense pump pulse generates a large density of excited carriers in the sample and a delayed weak test pulse probes the reflectivity or transmittivity in a broad spectral range, that can include also the excitonic resonances. The pump pulse will induce a coherent polarization in the material, which will undergo a rapid transition from coherent to incoherent regime as determined by dephasing processes. Owing to the above-bandgap excitation, the pump pulse photoexcites unbound charge carriers in the form of an electron-hole-plasma. Initially one expects a nonequilibrium hot-carrier distribution, which equilibrates due to carrier-carrier Coulomb scattering and cools down to the lattice temperature by means of carrier-phonon interaction. In the course of these processes, excited carriers will occupy the valleys of the band structure. \cite{steinhoff_nonequilibrium_2016,bertoni_generation_2016,selig_excitonic_2016,cadiz_excitonic_2017} Typically a quasi-equilibrium situation is established as the relaxation processes occur before recombination sets in\cite{palummo_exciton_2015,fang_control_2019}.
\\
The total density of excited electrons and holes $(e,h)$ can be calculated as
\begin{equation}
 \begin{split}
 n^{e/h}(t) = \frac{1}{\mathcal{A}}\sum \limits_{\bk} f_{\bk}^{e/h}(t) \, ,
 \end{split}
\label{eq:SBE_density}
\end{equation}
where $f_{\bk}^{e/h}(t) = \big\langle a^{\dagger,e/h}_\bk(t) \, a^{\vphantom{\dagger}e/h}_\bk(t) \big\rangle $ are the occupation probabilities of electron and hole Bloch states and $\mathcal{A}$ is the crystal area. 
The carrier dynamics is governed by the SBE, which are coupled equations of motion (EOM) for $f_{\bk}^{e/h}(t)$ and the microscopic inter-band polarization $\psi_{\bk}^{he}(t) = \big\langle a^{\vphantom{\dagger h}}_\bk(t) \, a^{\vphantom{\dagger}e}_\bk(t) \big\rangle $ driven by an electric field $\mathbf{E}(t)$ within the atomically thin layer. The field is given by the applied electromagnetic laser field $ \mathbf{E}_0(t)$ together with the reflected field reradiated by the induced coherent TMDC polarization, as determined by a solution of Maxwell's equations \cite{jahnke_linear_1997}:
\begin{eqnarray}
 \mathbf{E}(t)&=& \mathbf{E}_0(t)-\frac{\mu_0 c_0}{2} \frac{\partial}{\partial t} \mathbf{P}_{\textrm{TMDC}}(t) \nonumber \\
 &\approx& \mathbf{E}_0(t)+\frac{i\omega_0}{2\varepsilon_0 c_0 n} \mathbf{P}_{\textrm{TMDC}}(t) \, ,
\label{eq:effective_E}
\end{eqnarray}
where $\mathbf{P}_{\textrm{TMDC}}(t)= \frac{1}{\mathcal{A}} \sum_{\bk,h,e} \mathbf{d}_{\bk}^{he}  \psi_{\bk}^{he}(t) $ is the macroscopic polarization of the TMDC layer, $\omega_0$ is the pump laser frequency, $n$ is the refractive index of the dielectric environment and $\mathbf{d}_{\bk}^{he}$ denotes the dipole matrix element.
The approach takes into account the embedding of the TMDC into encapsulation layers. Specifically, $\mathbf{E}_0(t)$ is the externally applied optical field within the encapsulation layer.
The SBE are given by:
\begin{equation}
 \begin{split}
 \frac{\mathrm{d}}{\mathrm{d}t}f_{\bk}^{e/h}(t) = \frac{2}{\hbar} \mathrm{Im}\big[ (\Omega_{\bk}^{he})^* \psi_{\bk}^{he} \big]-\frac{f_{\bk}^{e/h}(t) - F_{\bk}^{e/h}}{\tau_{\textrm{relax}}}\,,
 \end{split}
\label{eq:SBE_population}
\end{equation}
\begin{eqnarray}
 i\hbar\frac{\mathrm{d}}{\mathrm{d}t}\psi_{\bk}^{he}(t) &=& \left(\tilde{\varepsilon}_{\bk}^{h}+\tilde{\varepsilon}_{\bk}^{e}-i\Gamma(n)\right)\psi_{\bk}^{he}(t) \nonumber \\
  &&-\left(1-f_{\bk}^{e}(t)-f_{\bk}^{h}(t)\right)\Omega_{\bk}^{he}  \, .
\label{eq:SBE_polarization}
\end{eqnarray}
Here, $\Omega_{\bk}^{he} = \mathbf{d}_{\bk}^{eh}\cdot \mathbf{E}(t)+\frac{1}{\mathcal{A}}\sum_{h',e',\bk'}W_{\bk\bk'\bk\bk'}^{eh'he'}\psi_{\bk'}^{h'e'}(t)$ is the Rabi energy renormalized by the screened Coulomb interaction between electrons and holes.
The inter-band Coulomb interaction give rise to excitonic resonances below the quasi-particle band gap as well as a redistribution of oscillator strength between band-to-band transitions. Screening of the Coulomb interaction consists of two fundamentally different contributions. “Background” screening due to virtual excitation of valence electrons into empty conduction-band states as well as the polarizability of surrounding dielectric material in a heterostructure is included via the dielectric function $\varepsilon^{-1}_{\textrm{b}}$ contained in Coulomb matrix elements $V_{\bk\bk'\bk\bk'}^{eh'he'}$. In addition to this, screening induced by photoexcited carriers is captured by the dielectric function $\varepsilon^{-1}_{\textrm{exc}}$, such that $W_{\bk\bk'\bk\bk'}^{eh'he'} = \varepsilon^{-1}_{\textrm{exc}}(|\bk-\bk'|)V_{\bk\bk'\bk\bk'}^{eh'he'}$ is the fully screened Coulomb matrix element.
The renormalized single-particle energies $\tilde{\varepsilon}_{\bk}^{\lambda}$ are discussed in the following section.
The Pauli exclusion principle enters via the blocking term $1-f_{\bk}^{e}(t)-f_{\bk}^{h}(t)$.
Excitation-power-dependent dephasing processes are described by $\Gamma(n)$.
To model the equilibration and cooling of excited carriers, we use the relaxation time approximation given by the last term of Eq.~(\ref{eq:SBE_population}). The non-equilibrium carrier population develops into a Fermi distribution $F_{\bk}^{e/h}$ for the respective species of charge carriers on a characteristic timescale $\tau_{\textrm{relax}}$.
As the carrier relaxation does not change the density of excited carriers, we use for $F_{\bk}^{e/h}$ a Fermi-Dirac function with the same carrier density as the momentary population $f_{\bk}^{e/h}(t)$. For the temperature of $F_{\bk}^{e/h}$ the lattice temperature is taken to model efficient carrier-phonon scattering. For the relaxation time due to carrier-carrier\cite{steinhoff_nonequilibrium_2016} and carrier-phonon\cite{selig_excitonic_2016,molina-sanchez_temperature-dependent_2016} scattering we use $\tau_{\textrm{relax}}=100$ fs.
\subsection{Coulomb interaction and band-structure renormalizations}\label{sec:bsren}
We combine the above theory of photoexcitation with band structures, dipole matrix elements and bare as well as screened Coulomb matrix elements on a G$_0$W$_0$ level as input for the SBE (\ref{eq:SBE_population})-(\ref{eq:SBE_polarization}). Our hybrid approach is designed to yield a numerically tractable and yet material-realistic model, using a minimal basis to represent band structures and interaction matrix elements on the same footing.
A six-band model (one valence band and two conduction bands for each spin degree of freedom) is considered to describe the photoexcitation physics around the fundamental band gap. We use a large set of momentum states spanning the full Brillouin zone, thereby including all band-structure valleys that can be occupied by excited electrons and holes, which directly influence the optical nonlinearities of the photoexcited carrier density. A second key ingredient is the realistic description of the long-range Coulomb interaction of excited carriers to accurately quantify exciton binding energies as well as band-gap renormalizations.
We utilize a lattice Hamiltonian $H_{\bk}^{\alpha\beta}$ formulated in a three-dimensional localized basis $\ket{\alpha}$
consisting of Wannier functions with dominant d-orbital character from the transition metal (Mo or W). The valence- and conduction-band splitting
caused by spin-orbit interaction is included in the lattice Hamiltonian along the lines of \cite{liu_three_2013,steinhoff_influence_2014}.
Diagonalization of the Hamiltonian yields the band structure $\varepsilon_{\bk}^{\lambda}$ and the Bloch states $\ket{\psi_{\bk}^{\lambda}}=\sum_{\alpha} c^{\lambda}_{\alpha,\bk}\ket{\bk,\alpha} $,
where the coefficients $c^{\lambda}_{\alpha,\bk}$ describe the momentum-dependent contribution of the orbital $\alpha$ to the Bloch band $\lambda$.
Further technical details on the approach are given in \cite{steinhoff_influence_2014}.
Using the so-called Peierls approach\cite{tomczak_optical_2009,steinhoff_influence_2014}, dipole matrix elements
can be calculated directly from the lattice Hamiltonian:
\begin{equation}
\bd^{\lambda\lambda'}_{\bk}=\frac{e}{i}\frac{1}{\varepsilon^{\lambda}_{\bk}-\varepsilon^{\lambda'}_{\bk}}\sum_{\alpha\beta}(c^{\lambda}_{\alpha,\bk})^*c^{\lambda'}_{\beta,\bk} \nabla_{\bk}H^{\alpha\beta}_{\bk}.
\label{eq:dipole_ME}
\end{equation}
Coulomb matrix elements for TMDC monolayers embedded in a dielectric environment are obtained from bare Coulomb matrix elements and the RPA dielectric function of a TMDC monolayer calculated in the localized basis $\ket{\alpha}$.\cite{steinhoff_exciton_2017} The bare matrix elements  and the dielectric function $\varepsilon_{\textrm{b}}$ including environmental screening effects are parametrized as a function of $|\bq|$ using the Wannier function continuum electrostatics approach \cite{rosner_wannier_2015}. This combines a macroscopic electrostatic model for the screening by the dielectric environment in a heterostructure with a localized description of Coulomb interaction and yields background-screened matrix elements $V^{\alpha\beta}_{|\bq|}$. The actual parametrization is provided in Ref.~\citenum{steinhoff_exciton_2017}. We obtain Coulomb matrix elements in the Bloch-state representation by a unitary transformation using the coefficients $c^{\lambda}_{\alpha,\bk}$:
\begin{equation}
\begin{split}
V^{\lambda_1 \lambda_2 \lambda_3 \lambda_4}_{\bk_1 \bk_2 \bk_3 \bk_4}& = \\
& \!\!\!\!\!\!\!\! \sum_{\alpha, \beta} \Big( c_{\alpha, \bk_1}^{\lambda_1} \Big)^{*} \Big( c_{\beta, \bk_2}^{\lambda_2} \Big)^{*} c_{\beta, \bk_3}^{\lambda_3} c_{\alpha, \bk_4}^{\lambda_4} V^{\alpha\beta}_{|\bk_1-\bk_4|} \, .
\end{split}
\label{eq:Coul_ME}
\end{equation}
The encapsulation of TMDC monolayers in hexagonal boron nitride (hBN) is frequently used to reduce inhomogeneous contributions to the linewidth, stemming from surface wrinkling or doping. \cite{cadiz_excitonic_2017} For such a situation we assume a dielectric environment for all four investigated TMDCs (\mos{}, \mose{}, \ws{}, \wse{}) with a dielectric constant of $\varepsilon_r = 4.5$.\cite{geick_normal_1966} In addition a narrow gap of $0.3$ nm between the monolayer and the surrounding hBN layers has been taken into account.\cite{florian_dielectric_2018} Due to the dielectric screening induced by hBN encapsulation, the single-particle band gap of the unexcited monolayer exhibits a shrinkage, that can be described in the $G\Delta W$-formalism as shown in Refs.~\citenum{florian_dielectric_2018,rohlfing_electronic_2010,winther_band_2017}. In the limit of static screening, the renormalization of the electron and hole bands $\lambda$ can be expressed as:
\begin{eqnarray}
\Sigma^{\lambda,\textrm{G}\Delta\textrm{W}}_{\bk} &=& \frac{1}{2} \frac{1}{\mathcal{A}}\sum\limits_{\bk'} \Delta V^{\textrm{G}\Delta\textrm{W},\lambda\lambda\lambda\lambda}_{\bk\bk'\bk\bk'} \, ,
\end{eqnarray}
where $\Delta V = V^{\textrm{HS}} - V^{\textrm{free}}$ is the difference of the Coulomb interaction macroscopically screened by the heterostructure and the freestanding monolayer.\cite{florian_dielectric_2018} The band structure of a TMDC monolayer in a dielectric environment is therefore given by
\begin{eqnarray}
\varepsilon_{\bk}^{\lambda,\textrm{G}\Delta\textrm{W}}=\varepsilon_{\bk}^{\lambda}+\Sigma^{\lambda,\textrm{G}\Delta\textrm{W}}_{\bk} \, .
\label{eq:energyrenorm}
\end{eqnarray}
As an example, we find a band-gap shrinkage of $360$ meV for an hBN/\mos{}/hBN heterostructure. 
\\
Our description focusses on photoexcitation of unbound electrons and holes on short time scales before a significant amount of excitons is formed, which involves slower acoustic-phonon-assisted relaxation processes. On these grounds, screening due to excited electrons and holes on top of the background screening is calculated using the plasma dielectric function $\varepsilon^{-1}_{\textrm{exc}}(|\bq|)$. In static approximation and in the long-wavelength limit,\cite{erben_excitation-induced_2018} the dielectric function assumes the well-known form $\varepsilon^{-1}_{\textrm{exc}}(|\bq|)=1+\kappa_{|\bq|}/|\bq|$.
\\
Under photoexcitation, the band structure experiences momentum-dependent renormalizations $\Sigma_{\bk}^{\lambda,\textrm{exc}}$ due to Coulomb interaction between photoexcited carriers:
\begin{eqnarray}
\tilde{\varepsilon}_{\bk}^{\lambda}=\varepsilon_{\bk}^{\lambda,\textrm{G}\Delta\textrm{W}}+\Sigma_{\bk}^{\lambda,\textrm{exc}} \, .
\label{eq:energyrenorm2}
\end{eqnarray}
With increasing excitation density, a pronounced shrinkage of the band gap is obtained in connection with the strong Coulomb interaction in TMDCs.\cite{ulstrup_ultrafast_2016,gao_renormalization_2017,liu_direct_2019,cunningham_photoinduced_2017,chernikov_population_2015, liu_direct_2019,fan_nonlinear_2017,xie_layer-modulated_2019,kang_universal_2017} Recently, it has been predicted that different shifts for the various band-structure valleys can lead to a transition from direct to indirect band gaps.\cite{erben_excitation-induced_2018} In general, band-structure renormalizations are composed of Hartree-Fock contributions and correlation terms.
Owing to the complexity of the used description in the full Brillouin zone, we rely on the screened exchange Coulomb hole (SXCH) approximation that uses a quasi-static approximation to correlation terms on a GW-level \cite{erben_excitation-induced_2018}.
The corresponding self-energy of excited carriers is given by
\begin{eqnarray}
\Sigma_{\bk}^{\lambda,\textrm{exc}} &=& \Sigma_{\bk}^{\lambda,\textrm{H}} + \Sigma_{\bk}^{\lambda,\textrm{U}} + \Sigma_{\bk}^{\lambda,\textrm{SX}} + \Sigma_{\bk}^{\lambda,\textrm{CH}}\, ,
\label{eq:selfenergy}
\end{eqnarray}
where $\Sigma_{\bk}^{\lambda,\textrm{H}}$ is the renormalization due to the Hartree interaction, $\Sigma_{\bk}^{\lambda,\textrm{U}}$ is the unscreened electron-hole exchange and $\Sigma_{\bk}^{\lambda,\textrm{SX}} + \Sigma_{\bk}^{\lambda,\textrm{CH}}$ constitutes the SXCH contribution. As discussed in Ref.~\citenum{erben_excitation-induced_2018}, $\Sigma_{\bk}^{\lambda,\textrm{U}}$ and $\Sigma_{\bk}^{\lambda,\textrm{SX}} + \Sigma_{\bk}^{\lambda,\textrm{CH}}$
induce the direct-indirect transition of the band gap. 
Since we take into account two conduction bands per spin, the self-energy (\ref{eq:selfenergy}) includes Coulomb interaction between electrons in different conduction bands $e$ and $e'$. Special care has to be taken when going beyond the Hartree-Fock level for these band combinations. In the SXCH scheme, the Hartree-Fock self-energy is augmented by including correlations on the GW level in static approximation. It is discussed in \cite{erben_excitation-induced_2018} that inter-band correlation terms on the GW level are sensitive to dynamical screening at roughly the energy difference between the involved bands.
Since photoexcited electrons reside at the K and $\Sigma$ points, where the energy difference between the conduction bands is large, relevant GW correlation terms would be sensitive to plasma screening at high frequencies on the order of $1$ eV. We assume that plasma screening, which is mainly caused by intra-band polarization, is weak at such frequencies, so that the inter-band correlations can be neglected. Hence self-energy terms connecting different conduction bands are treated on the Hartree-Fock level.
\subsection{Influence of excitation-induced dephasing}\label{sec:eid}
The optical absorption spectrum is directly influenced by the spectral width of excitonic and band-to-band transitions. The spectral HWHM is determined by the dephasing $\Gamma$ of inter-band polarizations entering the SBE (\ref{eq:SBE_polarization}). Due to the intrinsic connection between dephasing and carrier-carrier scattering, the dephasing rate is in general power-dependent and can therefore contribute to a nonlinear dependence of excited carrier density on pump fluence.
\begin{figure}[h!t]
\centering
\includegraphics[trim =5cm 10.5cm 5cm 10.5cm, clip, width=.5\textwidth]{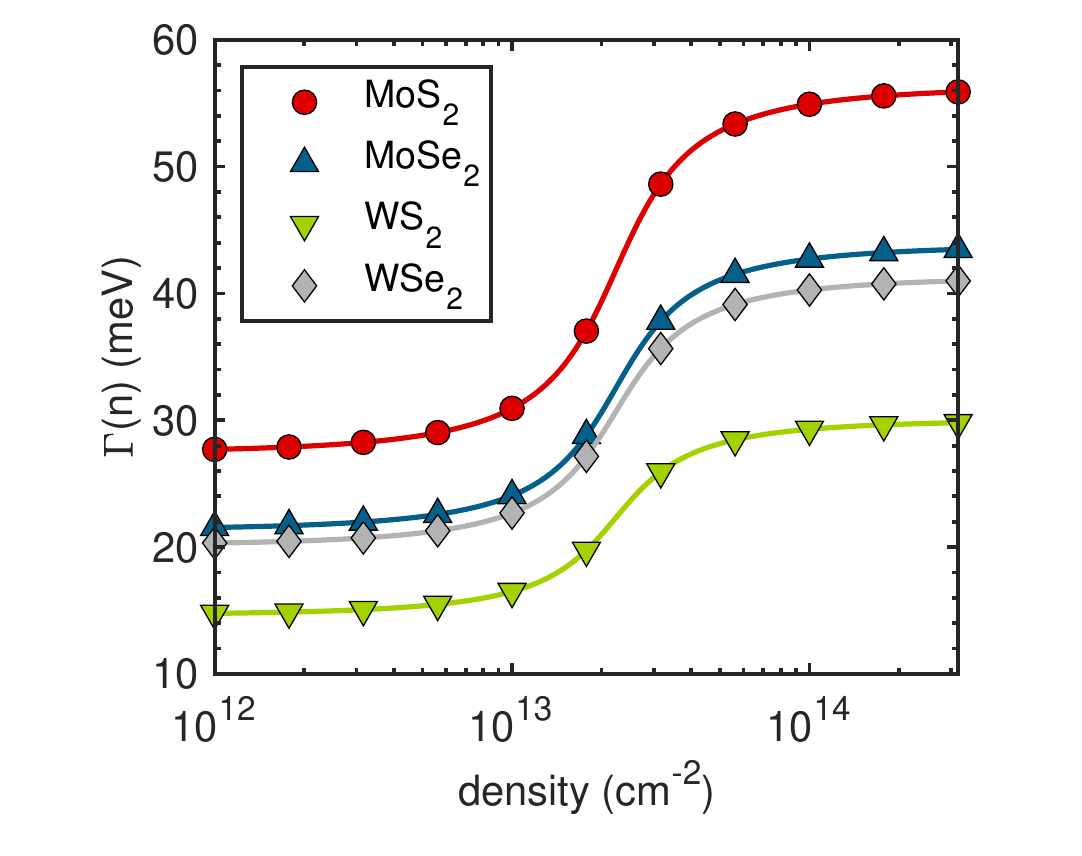}
\caption{Calculated carrier-density dependence of the excitation-induced dephasing for MoS$_2$, MoSe$_2$, WS$_2$, and WSe$_2$ obtained from a model that is described in detail in the text.
}
\label{fig:fig1}
\end{figure}
For the low-excitation regime (small pump fluence), the dominant contribution to dephasing stems from the coupling of charge carriers to phonons. 
This effect strongly depends on the lattice temperature and on the TMDC material. It has been quantified for the A exciton in Ref.~\cite{cadiz_excitonic_2017} and can be used to fix the value of the excitation-density dependent dephasing $\Gamma(n)$ at $n=0$.
To model the power-dependent contribution, we refer to Ref.\cite{sie_observation_2017}, where the increase of the A-exciton linewidth in \ws{} has been extracted from a pump-probe experiment for carrier densities up to several $10^{12}$ cm$^{-2}$. We assume that the Mott density, where excitons are fully dissociated into unbound carriers, marks a turning point in the density dependence of excitation-induced dephasing. Moreover, it can be expected that the dephasing rate saturates for large densities due to phase space filling in the carrier-carrier scattering rates.
These effects are phenomenologically captured by the expression
\begin{eqnarray}
\Gamma(n) &=& \left[ \arctan \left(\frac{ n - n_{\textrm{Mott}}}{\eta} - 1 \right) + \frac{\pi}{2}\right] \nonumber \\
&& \times \frac{\gamma_{0}(f-1)}{\pi}+\gamma_{0}\, ,
\end{eqnarray}
where $\gamma_0$ is taken from \cite{cadiz_excitonic_2017} to be $22.5$ meV for \mos{}, $17.5$ meV for \mose{}, $12$ meV for \ws{} and $16.5$ meV for \wse{}. The Mott density is approximated as $n_{\textrm{Mott}} = 1\times10^{13}$cm$^{-2}$\cite{steinhoff_exciton_2017}. The coefficients $\eta=1\times10^{13}$cm$^{-2}$ and $f=2.5$ can be estimated using the data from Ref.~\citenum{sie_observation_2017}. We assume $n_{\textrm{Mott}}$, $\eta$ and $f$ to be the same for all TMDC materials. For \mos{}, this yields a maximum dephasing rate of $56$ meV. The results for all considered materials are shown in Fig.~\ref{fig:fig1}.
\subsection{Pump pulse characteristics}
The fluence of the pump pulse propagating in a dielectric material with refractive index $n$ corresponds to the transmitted electromagnetic energy per area given by
\begin{equation}
 \begin{split}
  F=\int dt \left| \mathbf{S}(t) \right| = \varepsilon_0 c_0 n \int dt \left| \mathbf{E}_0(t) \right|^2\,, \end{split}
\label{eq:fluence}
\end{equation}
with $\textbf{S}$ denoting the Poynting vector. Here, $\mathbf{E}_0(t)$ is the pump laser field inside the hBN encapsulation layer with refractive index $n=\sqrt{\varepsilon_r}$. In specific experimental situations, the electric field inside the sample can be estimated from the field outside the sample, using a transfer matrix approach. We describe the laser field $\mathbf{E}_0(t)$ by a Gaussian pulse envelope with $150$ fs FWHM duration, circular light polarization, and incidence perpendicular to the TMD layer. We assume that the field intensity is given by the field intensity outside the sample, reduced by the reflectance $R=\left|\frac{1-n}{1+n}\right|^2$ of the outer boundary of the hBN capping layer. Throughout the paper, we refer to the fluence outside the sample that can be directly compared to experimental parameters.
\begin{suppinfo}
\end{suppinfo}
\begin{acknowledgement}
This work has been supported by the Deutsche Forschungsgemeinschaft through the graduate school Quantum Mechanic Materials Modelling (QM$^3$, RTG 2247) and project JA 619/18-1. We acknowledge a grant for CPU time at the HLRN (Göttingen/Berlin).
\end{acknowledgement}
\bibliography{DensityPaper_FINAL_SUB_A}

\providecommand{\latin}[1]{#1}
\makeatletter
\providecommand{\doi}
  {\begingroup\let\do\@makeother\dospecials
  \catcode`\{=1 \catcode`\}=2 \doi@aux}
\providecommand{\doi@aux}[1]{\endgroup\texttt{#1}}
\makeatother
\providecommand*\mcitethebibliography{\thebibliography}
\csname @ifundefined\endcsname{endmcitethebibliography}
  {\let\endmcitethebibliography\endthebibliography}{}
\begin{mcitethebibliography}{56}
\providecommand*\natexlab[1]{#1}
\providecommand*\mciteSetBstSublistMode[1]{}
\providecommand*\mciteSetBstMaxWidthForm[2]{}
\providecommand*\mciteBstWouldAddEndPuncttrue
  {\def\EndOfBibitem{\unskip.}}
\providecommand*\mciteBstWouldAddEndPunctfalse
  {\let\EndOfBibitem\relax}
\providecommand*\mciteSetBstMidEndSepPunct[3]{}
\providecommand*\mciteSetBstSublistLabelBeginEnd[3]{}
\providecommand*\EndOfBibitem{}
\mciteSetBstSublistMode{f}
\mciteSetBstMaxWidthForm{subitem}{(\alph{mcitesubitemcount})}
\mciteSetBstSublistLabelBeginEnd
  {\mcitemaxwidthsubitemform\space}
  {\relax}
  {\relax}

\bibitem[Xu \latin{et~al.}(2014)Xu, Yao, Xiao, and Heinz]{xu_spin_2014}
Xu,~X.; Yao,~W.; Xiao,~D.; Heinz,~T.~F. Spin and pseudospins in layered
  transition metal dichalcogenides. \emph{Nature Physics} \textbf{2014},
  \emph{10}, 343--350\relax
\mciteBstWouldAddEndPuncttrue
\mciteSetBstMidEndSepPunct{\mcitedefaultmidpunct}
{\mcitedefaultendpunct}{\mcitedefaultseppunct}\relax
\EndOfBibitem
\bibitem[Ugeda \latin{et~al.}(2014)Ugeda, Bradley, Shi, Jornada, Zhang, Qiu,
  Ruan, Mo, Hussain, Shen, Wang, Louie, and Crommie]{ugeda_giant_2014}
Ugeda,~M.~M.; Bradley,~A.~J.; Shi,~S.-F.; Jornada,~F. H.~d.; Zhang,~Y.;
  Qiu,~D.~Y.; Ruan,~W.; Mo,~S.-K.; Hussain,~Z.; Shen,~Z.-X.; Wang,~F.;
  Louie,~S.~G.; Crommie,~M.~F. Giant bandgap renormalization and excitonic
  effects in a monolayer transition metal dichalcogenide semiconductor.
  \emph{Nature Materials} \textbf{2014}, \emph{13}, 1091\relax
\mciteBstWouldAddEndPuncttrue
\mciteSetBstMidEndSepPunct{\mcitedefaultmidpunct}
{\mcitedefaultendpunct}{\mcitedefaultseppunct}\relax
\EndOfBibitem
\bibitem[Qiu \latin{et~al.}(2013)Qiu, da~Jornada, and Louie]{qiu_optical_2013}
Qiu,~D.~Y.; da~Jornada,~F.~H.; Louie,~S.~G. Optical Spectrum of
  {${\mathrm{MoS}}_{2}$:} Many-Body Effects and Diversity of Exciton States.
  \emph{Physical Review Letters} \textbf{2013}, \emph{111}, 216805\relax
\mciteBstWouldAddEndPuncttrue
\mciteSetBstMidEndSepPunct{\mcitedefaultmidpunct}
{\mcitedefaultendpunct}{\mcitedefaultseppunct}\relax
\EndOfBibitem
\bibitem[Chernikov \latin{et~al.}(2014)Chernikov, Berkelbach, Hill, Rigosi, Li,
  Aslan, Reichman, Hybertsen, and Heinz]{chernikov_exciton_2014}
Chernikov,~A.; Berkelbach,~T.~C.; Hill,~H.~M.; Rigosi,~A.; Li,~Y.;
  Aslan,~O.~B.; Reichman,~D.~R.; Hybertsen,~M.~S.; Heinz,~T.~F. Exciton Binding
  Energy and Nonhydrogenic Rydberg Series in Monolayer {${\mathrm{WS}}_{2}$}.
  \emph{Physical Review Letters} \textbf{2014}, \emph{113}, 076802\relax
\mciteBstWouldAddEndPuncttrue
\mciteSetBstMidEndSepPunct{\mcitedefaultmidpunct}
{\mcitedefaultendpunct}{\mcitedefaultseppunct}\relax
\EndOfBibitem
\bibitem[Ye \latin{et~al.}(2014)Ye, Cao, {O’Brien}, Zhu, Yin, Wang, Louie,
  and Zhang]{ye_probing_2014}
Ye,~Z.; Cao,~T.; {O’Brien},~K.; Zhu,~H.; Yin,~X.; Wang,~Y.; Louie,~S.~G.;
  Zhang,~X. Probing excitonic dark states in single-layer tungsten disulphide.
  \emph{Nature} \textbf{2014}, \emph{513}, 214--218\relax
\mciteBstWouldAddEndPuncttrue
\mciteSetBstMidEndSepPunct{\mcitedefaultmidpunct}
{\mcitedefaultendpunct}{\mcitedefaultseppunct}\relax
\EndOfBibitem
\bibitem[Tang \latin{et~al.}(2019)Tang, Mak, and Shan]{tang_long_2019}
Tang,~Y.; Mak,~K.~F.; Shan,~J. Long valley lifetime of dark excitons in
  single-layer {WSe} 2. \emph{Nature Communications} \textbf{2019}, \emph{10},
  1--7\relax
\mciteBstWouldAddEndPuncttrue
\mciteSetBstMidEndSepPunct{\mcitedefaultmidpunct}
{\mcitedefaultendpunct}{\mcitedefaultseppunct}\relax
\EndOfBibitem
\bibitem[Plechinger \latin{et~al.}(2016)Plechinger, Nagler, Arora, Schmidt,
  Chernikov, Águila, Christianen, Bratschitsch, Sch{\"u}ller, and
  Korn]{plechinger_trion_2016}
Plechinger,~G.; Nagler,~P.; Arora,~A.; Schmidt,~R.; Chernikov,~A.; Águila,~A.
  G.~d.; Christianen,~P. C.~M.; Bratschitsch,~R.; Sch{\"u}ller,~C.; Korn,~T.
  Trion fine structure and coupled spin–valley dynamics in monolayer tungsten
  disulfide. \emph{Nature Communications} \textbf{2016}, \emph{7}, 1--9\relax
\mciteBstWouldAddEndPuncttrue
\mciteSetBstMidEndSepPunct{\mcitedefaultmidpunct}
{\mcitedefaultendpunct}{\mcitedefaultseppunct}\relax
\EndOfBibitem
\bibitem[Florian \latin{et~al.}(2018)Florian, Hartmann, Steinhoff, Klein,
  Holleitner, Finley, Wehling, Kaniber, and Gies]{florian_dielectric_2018}
Florian,~M.; Hartmann,~M.; Steinhoff,~A.; Klein,~J.; Holleitner,~A.~W.;
  Finley,~J.~J.; Wehling,~T.~O.; Kaniber,~M.; Gies,~C. The Dielectric Impact of
  Layer Distances on Exciton and Trion Binding Energies in van der Waals
  Heterostructures. \emph{Nano Letters} \textbf{2018}, \emph{18},
  2725--2732\relax
\mciteBstWouldAddEndPuncttrue
\mciteSetBstMidEndSepPunct{\mcitedefaultmidpunct}
{\mcitedefaultendpunct}{\mcitedefaultseppunct}\relax
\EndOfBibitem
\bibitem[Hao \latin{et~al.}(2017)Hao, Specht, Nagler, Xu, Tran, Singh, Dass,
  Sch{\"u}ller, Korn, Richter, Knorr, Li, and Moody]{hao_neutral_2017}
Hao,~K.; Specht,~J.~F.; Nagler,~P.; Xu,~L.; Tran,~K.; Singh,~A.; Dass,~C.~K.;
  Sch{\"u}ller,~C.; Korn,~T.; Richter,~M.; Knorr,~A.; Li,~X.; Moody,~G. Neutral
  and charged inter-valley biexcitons in monolayer {MoSe} 2. \emph{Nature
  Communications} \textbf{2017}, \emph{8}, 1--7\relax
\mciteBstWouldAddEndPuncttrue
\mciteSetBstMidEndSepPunct{\mcitedefaultmidpunct}
{\mcitedefaultendpunct}{\mcitedefaultseppunct}\relax
\EndOfBibitem
\bibitem[Steinhoff \latin{et~al.}(2018)Steinhoff, Florian, Singh, Tran,
  Kolarczik, Helmrich, Achtstein, Woggon, Owschimikow, Jahnke, and
  Li]{steinhoff_biexciton_2018}
Steinhoff,~A.; Florian,~M.; Singh,~A.; Tran,~K.; Kolarczik,~M.; Helmrich,~S.;
  Achtstein,~A.~W.; Woggon,~U.; Owschimikow,~N.; Jahnke,~F.; Li,~X. Biexciton
  fine structure in monolayer transition metal dichalcogenides. \emph{Nature
  Physics} \textbf{2018}, \emph{14}, 1199--1204\relax
\mciteBstWouldAddEndPuncttrue
\mciteSetBstMidEndSepPunct{\mcitedefaultmidpunct}
{\mcitedefaultendpunct}{\mcitedefaultseppunct}\relax
\EndOfBibitem
\bibitem[Steinhoff \latin{et~al.}(2014)Steinhoff, R{\"o}sner, Jahnke, Wehling,
  and Gies]{steinhoff_influence_2014}
Steinhoff,~A.; R{\"o}sner,~M.; Jahnke,~F.; Wehling,~T.~O.; Gies,~C. Influence
  of Excited Carriers on the Optical and Electronic Properties of {MoS$_2$}.
  \emph{Nano Letters} \textbf{2014}, \emph{14}, 3743--3748\relax
\mciteBstWouldAddEndPuncttrue
\mciteSetBstMidEndSepPunct{\mcitedefaultmidpunct}
{\mcitedefaultendpunct}{\mcitedefaultseppunct}\relax
\EndOfBibitem
\bibitem[Meckbach \latin{et~al.}(2018)Meckbach, Stroucken, and
  Koch]{meckbach_giant_2018}
Meckbach,~L.; Stroucken,~T.; Koch,~S.~W. Giant excitation induced bandgap
  renormalization in TMDC monolayers. \emph{Applied Physics Letters}
  \textbf{2018}, \emph{112}, 061104\relax
\mciteBstWouldAddEndPuncttrue
\mciteSetBstMidEndSepPunct{\mcitedefaultmidpunct}
{\mcitedefaultendpunct}{\mcitedefaultseppunct}\relax
\EndOfBibitem
\bibitem[Kaasbjerg \latin{et~al.}(2014)Kaasbjerg, Bhargavi, and
  Kubakaddi]{kaasbjerg_hot-electron_2014}
Kaasbjerg,~K.; Bhargavi,~K.~S.; Kubakaddi,~S.~S. Hot-electron cooling by
  acoustic and optical phonons in monolayers of ${\mathrm{MoS}}_{2}$ and other
  transition-metal dichalcogenides. \emph{Phys. Rev. B} \textbf{2014},
  \emph{90}, 165436\relax
\mciteBstWouldAddEndPuncttrue
\mciteSetBstMidEndSepPunct{\mcitedefaultmidpunct}
{\mcitedefaultendpunct}{\mcitedefaultseppunct}\relax
\EndOfBibitem
\bibitem[Steinhoff \latin{et~al.}(2016)Steinhoff, Florian, R{\"o}sner, Lorke,
  Wehling, Gies, and Jahnke]{steinhoff_nonequilibrium_2016}
Steinhoff,~A.; Florian,~M.; R{\"o}sner,~M.; Lorke,~M.; Wehling,~T.~O.;
  Gies,~C.; Jahnke,~F. Nonequilibrium carrier dynamics in transition metal
  dichalcogenide semiconductors. \emph{{2D} Materials} \textbf{2016}, \emph{3},
  031006\relax
\mciteBstWouldAddEndPuncttrue
\mciteSetBstMidEndSepPunct{\mcitedefaultmidpunct}
{\mcitedefaultendpunct}{\mcitedefaultseppunct}\relax
\EndOfBibitem
\bibitem[{Danovich} \latin{et~al.}(2017){Danovich}, {Aleiner}, {Drummond}, and
  {Fal’ko}]{danovich_fast_2017}
{Danovich},~M.; {Aleiner},~I.~L.; {Drummond},~N.~D.; {Fal’ko},~V.~I. Fast
  Relaxation of Photo-Excited Carriers in 2-D Transition Metal Dichalcogenides.
  \emph{IEEE Journal of Selected Topics in Quantum Electronics} \textbf{2017},
  \emph{23}, 168--172\relax
\mciteBstWouldAddEndPuncttrue
\mciteSetBstMidEndSepPunct{\mcitedefaultmidpunct}
{\mcitedefaultendpunct}{\mcitedefaultseppunct}\relax
\EndOfBibitem
\bibitem[Steinhoff \latin{et~al.}(2017)Steinhoff, Florian, R{\"o}sner,
  Sch{\"o}nhoff, Wehling, and Jahnke]{steinhoff_exciton_2017}
Steinhoff,~A.; Florian,~M.; R{\"o}sner,~M.; Sch{\"o}nhoff,~G.; Wehling,~T.~O.;
  Jahnke,~F. Exciton fission in monolayer transition metal dichalcogenide
  semiconductors. \emph{Nature Communications} \textbf{2017}, \emph{8},
  1166\relax
\mciteBstWouldAddEndPuncttrue
\mciteSetBstMidEndSepPunct{\mcitedefaultmidpunct}
{\mcitedefaultendpunct}{\mcitedefaultseppunct}\relax
\EndOfBibitem
\bibitem[Pospischil \latin{et~al.}(2014)Pospischil, Furchi, and
  Mueller]{pospischil_solar-energy_2014}
Pospischil,~A.; Furchi,~M.~M.; Mueller,~T. Solar-energy conversion and light
  emission in an atomic monolayer p–n diode. \emph{Nature Nanotechnology}
  \textbf{2014}, \emph{9}, 257--261\relax
\mciteBstWouldAddEndPuncttrue
\mciteSetBstMidEndSepPunct{\mcitedefaultmidpunct}
{\mcitedefaultendpunct}{\mcitedefaultseppunct}\relax
\EndOfBibitem
\bibitem[Baugher \latin{et~al.}(2014)Baugher, Churchill, Yang, and
  Jarillo-Herrero]{baugher_optoelectronic_2014}
Baugher,~B. W.~H.; Churchill,~H. O.~H.; Yang,~Y.; Jarillo-Herrero,~P.
  Optoelectronic devices based on electrically tunable p–n diodes in a
  monolayer dichalcogenide. \emph{Nature Nanotechnology} \textbf{2014},
  \emph{9}, 262--267\relax
\mciteBstWouldAddEndPuncttrue
\mciteSetBstMidEndSepPunct{\mcitedefaultmidpunct}
{\mcitedefaultendpunct}{\mcitedefaultseppunct}\relax
\EndOfBibitem
\bibitem[Ross \latin{et~al.}(2014)Ross, Klement, Jones, Ghimire, Yan, Mandrus,
  Taniguchi, Watanabe, Kitamura, Yao, Cobden, and Xu]{ross_electrically_2014}
Ross,~J.~S.; Klement,~P.; Jones,~A.~M.; Ghimire,~N.~J.; Yan,~J.;
  Mandrus,~D.~G.; Taniguchi,~T.; Watanabe,~K.; Kitamura,~K.; Yao,~W.;
  Cobden,~D.~H.; Xu,~X. Electrically tunable excitonic light-emitting diodes
  based on monolayer {WSe$_2$} p–n junctions. \emph{Nature Nanotechnology}
  \textbf{2014}, \emph{9}, 268--272\relax
\mciteBstWouldAddEndPuncttrue
\mciteSetBstMidEndSepPunct{\mcitedefaultmidpunct}
{\mcitedefaultendpunct}{\mcitedefaultseppunct}\relax
\EndOfBibitem
\bibitem[Withers \latin{et~al.}(2015)Withers, Pozo-Zamudio, Mishchenko, Rooney,
  Gholinia, Watanabe, Taniguchi, Haigh, Geim, Tartakovskii, and
  Novoselov]{withers_light-emitting_2015}
Withers,~F.; Pozo-Zamudio,~O.~D.; Mishchenko,~A.; Rooney,~A.~P.; Gholinia,~A.;
  Watanabe,~K.; Taniguchi,~T.; Haigh,~S.~J.; Geim,~A.~K.; Tartakovskii,~A.~I.;
  Novoselov,~K.~S. Light-emitting diodes by band-structure engineering in van
  der Waals heterostructures. \emph{Nature Materials} \textbf{2015}, \emph{14},
  301--306\relax
\mciteBstWouldAddEndPuncttrue
\mciteSetBstMidEndSepPunct{\mcitedefaultmidpunct}
{\mcitedefaultendpunct}{\mcitedefaultseppunct}\relax
\EndOfBibitem
\bibitem[Wu \latin{et~al.}(2015)Wu, Buckley, Schaibley, Feng, Yan, Mandrus,
  Hatami, Yao, Vučković, Majumdar, and Xu]{wu_monolayer_2015}
Wu,~S.; Buckley,~S.; Schaibley,~J.~R.; Feng,~L.; Yan,~J.; Mandrus,~D.~G.;
  Hatami,~F.; Yao,~W.; Vučković,~J.; Majumdar,~A.; Xu,~X. Monolayer
  semiconductor nanocavity lasers with ultralow thresholds. \emph{Nature}
  \textbf{2015}, \emph{520}, 69--72\relax
\mciteBstWouldAddEndPuncttrue
\mciteSetBstMidEndSepPunct{\mcitedefaultmidpunct}
{\mcitedefaultendpunct}{\mcitedefaultseppunct}\relax
\EndOfBibitem
\bibitem[Ye \latin{et~al.}(2015)Ye, Wong, Lu, Ni, Zhu, Chen, Wang, and
  Zhang]{ye_monolayer_2015}
Ye,~Y.; Wong,~Z.~J.; Lu,~X.; Ni,~X.; Zhu,~H.; Chen,~X.; Wang,~Y.; Zhang,~X.
  Monolayer excitonic laser. \emph{Nature Photonics} \textbf{2015}, \emph{9},
  733--737\relax
\mciteBstWouldAddEndPuncttrue
\mciteSetBstMidEndSepPunct{\mcitedefaultmidpunct}
{\mcitedefaultendpunct}{\mcitedefaultseppunct}\relax
\EndOfBibitem
\bibitem[Salehzadeh \latin{et~al.}(2015)Salehzadeh, Djavid, Tran, Shih, and
  Mi]{salehzadeh_optically_2015}
Salehzadeh,~O.; Djavid,~M.; Tran,~N.~H.; Shih,~I.; Mi,~Z. Optically Pumped
  Two-Dimensional {MoS$_2$} Lasers Operating at Room-Temperature. \emph{Nano
  Letters} \textbf{2015}, \emph{15}, 5302--5306\relax
\mciteBstWouldAddEndPuncttrue
\mciteSetBstMidEndSepPunct{\mcitedefaultmidpunct}
{\mcitedefaultendpunct}{\mcitedefaultseppunct}\relax
\EndOfBibitem
\bibitem[Li \latin{et~al.}(2017)Li, Zhang, Huang, Sun, Fan, Feng, Wang, and
  Ning]{li_room-temperature_2017}
Li,~Y.; Zhang,~J.; Huang,~D.; Sun,~H.; Fan,~F.; Feng,~J.; Wang,~Z.; Ning,~C.~Z.
  Room-temperature continuous-wave lasing from monolayer molybdenum ditelluride
  integrated with a silicon nanobeam cavity. \emph{Nature Nanotechnology}
  \textbf{2017}, \emph{12}, 987--992\relax
\mciteBstWouldAddEndPuncttrue
\mciteSetBstMidEndSepPunct{\mcitedefaultmidpunct}
{\mcitedefaultendpunct}{\mcitedefaultseppunct}\relax
\EndOfBibitem
\bibitem[Lohof \latin{et~al.}(2019)Lohof, Steinhoff, Florian, Lorke, Erben,
  Jahnke, and Gies]{lohof_prospects_2019}
Lohof,~F.; Steinhoff,~A.; Florian,~M.; Lorke,~M.; Erben,~D.; Jahnke,~F.;
  Gies,~C. Prospects and Limitations of Transition Metal Dichalcogenide Laser
  Gain Materials. \emph{Nano Letters} \textbf{2019}, \emph{19}, 210--217\relax
\mciteBstWouldAddEndPuncttrue
\mciteSetBstMidEndSepPunct{\mcitedefaultmidpunct}
{\mcitedefaultendpunct}{\mcitedefaultseppunct}\relax
\EndOfBibitem
\bibitem[He \latin{et~al.}(2014)He, Kumar, Zhao, Wang, Mak, Zhao, and
  Shan]{he_tightly_2014}
He,~K.; Kumar,~N.; Zhao,~L.; Wang,~Z.; Mak,~K.~F.; Zhao,~H.; Shan,~J. Tightly
  Bound Excitons in Monolayer {${\mathrm{WSe}}_{2}$}. \emph{Physical Review
  Letters} \textbf{2014}, \emph{113}, 026803\relax
\mciteBstWouldAddEndPuncttrue
\mciteSetBstMidEndSepPunct{\mcitedefaultmidpunct}
{\mcitedefaultendpunct}{\mcitedefaultseppunct}\relax
\EndOfBibitem
\bibitem[Steinhoff \latin{et~al.}(2015)Steinhoff, Kim, Jahnke, R{\"o}sner, Kim,
  Lee, Han, Jeong, Wehling, and Gies]{steinhoff_efficient_2015}
Steinhoff,~A.; Kim,~J.-H.; Jahnke,~F.; R{\"o}sner,~M.; Kim,~D.-S.; Lee,~C.;
  Han,~G.~H.; Jeong,~M.~S.; Wehling,~T.~O.; Gies,~C. Efficient Excitonic
  Photoluminescence in Direct and Indirect Band Gap Monolayer {MoS$_2$}.
  \emph{Nano Letters} \textbf{2015}, \emph{15}, 6841--6847\relax
\mciteBstWouldAddEndPuncttrue
\mciteSetBstMidEndSepPunct{\mcitedefaultmidpunct}
{\mcitedefaultendpunct}{\mcitedefaultseppunct}\relax
\EndOfBibitem
\bibitem[Klots \latin{et~al.}(2014)Klots, Newaz, Wang, Prasai, Krzyzanowska,
  Lin, Caudel, Ghimire, Yan, Ivanov, Velizhanin, Burger, Mandrus, Tolk,
  Pantelides, and Bolotin]{klots_probing_2014}
Klots,~A.~R.; Newaz,~A. K.~M.; Wang,~B.; Prasai,~D.; Krzyzanowska,~H.; Lin,~J.;
  Caudel,~D.; Ghimire,~N.~J.; Yan,~J.; Ivanov,~B.~L.; Velizhanin,~K.~A.;
  Burger,~A.; Mandrus,~D.~G.; Tolk,~N.~H.; Pantelides,~S.~T.; Bolotin,~K.~I.
  Probing excitonic states in suspended two-dimensional semiconductors by
  photocurrent spectroscopy. \emph{Scientific Reports} \textbf{2014}, \emph{4},
  6608\relax
\mciteBstWouldAddEndPuncttrue
\mciteSetBstMidEndSepPunct{\mcitedefaultmidpunct}
{\mcitedefaultendpunct}{\mcitedefaultseppunct}\relax
\EndOfBibitem
\bibitem[Chernikov \latin{et~al.}(2015)Chernikov, Ruppert, Hill, Rigosi, and
  Heinz]{chernikov_population_2015}
Chernikov,~A.; Ruppert,~C.; Hill,~H.~M.; Rigosi,~A.~F.; Heinz,~T.~F. Population
  inversion and giant bandgap renormalization in atomically thin {WS$_2$}
  layers. \emph{Nature Photonics} \textbf{2015}, \emph{9}, 466\relax
\mciteBstWouldAddEndPuncttrue
\mciteSetBstMidEndSepPunct{\mcitedefaultmidpunct}
{\mcitedefaultendpunct}{\mcitedefaultseppunct}\relax
\EndOfBibitem
\bibitem[Wang \latin{et~al.}(2019)Wang, Ardelean, Bai, Steinhoff, Florian,
  Jahnke, Xu, Kira, Hone, and Zhu]{wang_optical_2019}
Wang,~J.; Ardelean,~J.; Bai,~Y.; Steinhoff,~A.; Florian,~M.; Jahnke,~F.;
  Xu,~X.; Kira,~M.; Hone,~J.; Zhu,~X.-Y. Optical generation of high carrier
  densities in {2D} semiconductor heterobilayers. \emph{Science Advances}
  \textbf{2019}, \emph{5}, eaax0145\relax
\mciteBstWouldAddEndPuncttrue
\mciteSetBstMidEndSepPunct{\mcitedefaultmidpunct}
{\mcitedefaultendpunct}{\mcitedefaultseppunct}\relax
\EndOfBibitem
\bibitem[Sie \latin{et~al.}(2017)Sie, Steinhoff, Gies, Lui, Ma, R{\"o}sner,
  Sch{\"o}nhoff, Jahnke, Wehling, Lee, Kong, Jarillo-Herrero, and
  Gedik]{sie_observation_2017}
Sie,~E.~J.; Steinhoff,~A.; Gies,~C.; Lui,~C.~H.; Ma,~Q.; R{\"o}sner,~M.;
  Sch{\"o}nhoff,~G.; Jahnke,~F.; Wehling,~T.~O.; Lee,~Y.-H.; Kong,~J.;
  Jarillo-Herrero,~P.; Gedik,~N. Observation of Exciton {Redshift–Blueshift}
  Crossover in Monolayer {WS$_2$}. \emph{Nano Letters} \textbf{2017},
  \emph{17}, 4210--4216\relax
\mciteBstWouldAddEndPuncttrue
\mciteSetBstMidEndSepPunct{\mcitedefaultmidpunct}
{\mcitedefaultendpunct}{\mcitedefaultseppunct}\relax
\EndOfBibitem
\bibitem[Bataller \latin{et~al.}(2019)Bataller, Younts, Rustagi, Yu, Ardekani,
  Kemper, Cao, and Gundogdu]{bataller_dense_2019}
Bataller,~A.~W.; Younts,~R.~A.; Rustagi,~A.; Yu,~Y.; Ardekani,~H.; Kemper,~A.;
  Cao,~L.; Gundogdu,~K. Dense {Electron–Hole} Plasma Formation and Ultralong
  Charge Lifetime in Monolayer {MoS$_2$} via Material Tuning. \emph{Nano
  Letters} \textbf{2019}, \emph{19}, 1104--1111\relax
\mciteBstWouldAddEndPuncttrue
\mciteSetBstMidEndSepPunct{\mcitedefaultmidpunct}
{\mcitedefaultendpunct}{\mcitedefaultseppunct}\relax
\EndOfBibitem
\bibitem[Li \latin{et~al.}(2019)Li, Shi, Chen, Mi, Du, Sui, Jiang, Liu, Xu, and
  Liu]{li_slow_2019}
Li,~Y.; Shi,~J.; Chen,~H.; Mi,~Y.; Du,~W.; Sui,~X.; Jiang,~C.; Liu,~W.; Xu,~H.;
  Liu,~X. Slow Cooling of High-Energy C Excitons Is Limited by
  Intervalley-Transfer in Monolayer {MoS2}. \emph{Laser \& Photonics Reviews}
  \textbf{2019}, \emph{13}, 1800270\relax
\mciteBstWouldAddEndPuncttrue
\mciteSetBstMidEndSepPunct{\mcitedefaultmidpunct}
{\mcitedefaultendpunct}{\mcitedefaultseppunct}\relax
\EndOfBibitem
\bibitem[Haug and Koch(2004)Haug, and Koch]{haug_quantum_2004}
Haug,~H.; Koch,~S. \emph{Quantum theory of the optical and electronic
  properties of semiconductors}; World Scientific, 2004\relax
\mciteBstWouldAddEndPuncttrue
\mciteSetBstMidEndSepPunct{\mcitedefaultmidpunct}
{\mcitedefaultendpunct}{\mcitedefaultseppunct}\relax
\EndOfBibitem
\bibitem[Jahnke \latin{et~al.}(1997)Jahnke, Kira, and Koch]{jahnke_linear_1997}
Jahnke,~F.; Kira,~M.; Koch,~S. Linear and nonlinear optical properties of
  excitons in semiconductor quantum wells and microcavities. \emph{Z. Phys. B}
  \textbf{1997}, \emph{104}, 559--572\relax
\mciteBstWouldAddEndPuncttrue
\mciteSetBstMidEndSepPunct{\mcitedefaultmidpunct}
{\mcitedefaultendpunct}{\mcitedefaultseppunct}\relax
\EndOfBibitem
\bibitem[Erben \latin{et~al.}(2018)Erben, Steinhoff, Gies, Schönhoff, Wehling,
  and Jahnke]{erben_excitation-induced_2018}
Erben,~D.; Steinhoff,~A.; Gies,~C.; Schönhoff,~G.; Wehling,~T.~O.; Jahnke,~F.
  Excitation-induced transition to indirect band gaps in atomically thin
  transition-metal dichalcogenide semiconductors. \emph{Physical Review B}
  \textbf{2018}, \emph{98}, 035434\relax
\mciteBstWouldAddEndPuncttrue
\mciteSetBstMidEndSepPunct{\mcitedefaultmidpunct}
{\mcitedefaultendpunct}{\mcitedefaultseppunct}\relax
\EndOfBibitem
\bibitem[Fang \latin{et~al.}(2019)Fang, Han, Robert, Semina, Lagarde, Courtade,
  Taniguchi, Watanabe, Amand, Urbaszek, Glazov, and Marie]{fang_control_2019}
Fang,~H.~H.; Han,~B.; Robert,~C.; Semina,~M.~A.; Lagarde,~D.; Courtade,~E.;
  Taniguchi,~T.; Watanabe,~K.; Amand,~T.; Urbaszek,~B.; Glazov,~M.~M.;
  Marie,~X. Control of the Exciton Radiative Lifetime in van der Waals
  Heterostructures. \textbf{2019}, \relax
\mciteBstWouldAddEndPunctfalse
\mciteSetBstMidEndSepPunct{\mcitedefaultmidpunct}
{}{\mcitedefaultseppunct}\relax
\EndOfBibitem
\bibitem[Palummo \latin{et~al.}(2015)Palummo, Bernardi, and
  Grossman]{palummo_exciton_2015}
Palummo,~M.; Bernardi,~M.; Grossman,~J.~C. Exciton Radiative Lifetimes in
  Two-Dimensional Transition Metal Dichalcogenides. \emph{Nano Letters}
  \textbf{2015}, \emph{15}, 2794--2800\relax
\mciteBstWouldAddEndPuncttrue
\mciteSetBstMidEndSepPunct{\mcitedefaultmidpunct}
{\mcitedefaultendpunct}{\mcitedefaultseppunct}\relax
\EndOfBibitem
\bibitem[Bertoni \latin{et~al.}(2016)Bertoni, Nicholson, Waldecker,
  H{\"u}bener, Monney, De~Giovannini, Puppin, Hoesch, Springate, Chapman,
  Cacho, Wolf, Rubio, and Ernstorfer]{bertoni_generation_2016}
Bertoni,~R.; Nicholson,~C.; Waldecker,~L.; H{\"u}bener,~H.; Monney,~C.;
  De~Giovannini,~U.; Puppin,~M.; Hoesch,~M.; Springate,~E.; Chapman,~R.;
  Cacho,~C.; Wolf,~M.; Rubio,~A.; Ernstorfer,~R. Generation and Evolution of
  Spin-, Valley-, and Layer-Polarized Excited Carriers in Inversion-Symmetric
  {${\mathrm{WSe}}_{2}$}. \emph{Physical Review Letters} \textbf{2016},
  \emph{117}, 277201\relax
\mciteBstWouldAddEndPuncttrue
\mciteSetBstMidEndSepPunct{\mcitedefaultmidpunct}
{\mcitedefaultendpunct}{\mcitedefaultseppunct}\relax
\EndOfBibitem
\bibitem[Selig \latin{et~al.}(2016)Selig, Bergh{\"a}user, Raja, Nagler,
  Sch{\"u}ller, Heinz, Korn, Chernikov, Malic, and Knorr]{selig_excitonic_2016}
Selig,~M.; Bergh{\"a}user,~G.; Raja,~A.; Nagler,~P.; Sch{\"u}ller,~C.;
  Heinz,~T.~F.; Korn,~T.; Chernikov,~A.; Malic,~E.; Knorr,~A. Excitonic
  linewidth and coherence lifetime in monolayer transition metal
  dichalcogenides. \emph{Nature Communications} \textbf{2016}, \emph{7},
  1--6\relax
\mciteBstWouldAddEndPuncttrue
\mciteSetBstMidEndSepPunct{\mcitedefaultmidpunct}
{\mcitedefaultendpunct}{\mcitedefaultseppunct}\relax
\EndOfBibitem
\bibitem[Cadiz \latin{et~al.}(2017)Cadiz, Courtade, Robert, Wang, Shen, Cai,
  Taniguchi, Watanabe, Carrere, Lagarde, Manca, Amand, Renucci, Tongay, Marie,
  and Urbaszek]{cadiz_excitonic_2017}
Cadiz,~F.; Courtade,~E.; Robert,~C.; Wang,~G.; Shen,~Y.; Cai,~H.;
  Taniguchi,~T.; Watanabe,~K.; Carrere,~H.; Lagarde,~D.; Manca,~M.; Amand,~T.;
  Renucci,~P.; Tongay,~S.; Marie,~X.; Urbaszek,~B. Excitonic Linewidth
  Approaching the Homogeneous Limit in {${\mathrm{MoS}}_{2}$-Based} van der
  Waals Heterostructures. \emph{Physical Review X} \textbf{2017}, \emph{7},
  021026\relax
\mciteBstWouldAddEndPuncttrue
\mciteSetBstMidEndSepPunct{\mcitedefaultmidpunct}
{\mcitedefaultendpunct}{\mcitedefaultseppunct}\relax
\EndOfBibitem
\bibitem[Molina-Sánchez \latin{et~al.}(2016)Molina-Sánchez, Palummo, Marini,
  and Wirtz]{molina-sanchez_temperature-dependent_2016}
Molina-Sánchez,~A.; Palummo,~M.; Marini,~A.; Wirtz,~L. Temperature-dependent
  excitonic effects in the optical properties of single-layer
  {${\mathrm{MoS}}_{2}$}. \emph{Physical Review B} \textbf{2016}, \emph{93},
  155435\relax
\mciteBstWouldAddEndPuncttrue
\mciteSetBstMidEndSepPunct{\mcitedefaultmidpunct}
{\mcitedefaultendpunct}{\mcitedefaultseppunct}\relax
\EndOfBibitem
\bibitem[Liu \latin{et~al.}(2013)Liu, Shan, Yao, Yao, and Xiao]{liu_three_2013}
Liu,~G.-B.; Shan,~W.-Y.; Yao,~Y.; Yao,~W.; Xiao,~D. Three-band tight-binding
  model for monolayers of group-VIB transition metal dichalcogenides.
  \emph{Phys. Rev. B} \textbf{2013}, \emph{88}, 085433\relax
\mciteBstWouldAddEndPuncttrue
\mciteSetBstMidEndSepPunct{\mcitedefaultmidpunct}
{\mcitedefaultendpunct}{\mcitedefaultseppunct}\relax
\EndOfBibitem
\bibitem[Tomczak and Biermann(2009)Tomczak, and Biermann]{tomczak_optical_2009}
Tomczak,~J.~M.; Biermann,~S. Optical properties of correlated materials:
  Generalized Peierls approach and its application to ${\text{VO}}_{2}$.
  \emph{Phys. Rev. B} \textbf{2009}, \emph{80}, 085117\relax
\mciteBstWouldAddEndPuncttrue
\mciteSetBstMidEndSepPunct{\mcitedefaultmidpunct}
{\mcitedefaultendpunct}{\mcitedefaultseppunct}\relax
\EndOfBibitem
\bibitem[R{\"o}sner \latin{et~al.}(2015)R{\"o}sner, Şaşıoğlu, Friedrich,
  Bl{\"u}gel, and Wehling]{rosner_wannier_2015}
R{\"o}sner,~M.; Şaşıoğlu,~E.; Friedrich,~C.; Bl{\"u}gel,~S.; Wehling,~T.~O.
  Wannier function approach to realistic Coulomb interactions in layered
  materials and heterostructures. \emph{Physical Review B} \textbf{2015},
  \emph{92}, 085102\relax
\mciteBstWouldAddEndPuncttrue
\mciteSetBstMidEndSepPunct{\mcitedefaultmidpunct}
{\mcitedefaultendpunct}{\mcitedefaultseppunct}\relax
\EndOfBibitem
\bibitem[Geick \latin{et~al.}(1966)Geick, Perry, and
  Rupprecht]{geick_normal_1966}
Geick,~R.; Perry,~C.~H.; Rupprecht,~G. Normal Modes in Hexagonal Boron Nitride.
  \emph{Phys. Rev.} \textbf{1966}, \emph{146}, 543--547\relax
\mciteBstWouldAddEndPuncttrue
\mciteSetBstMidEndSepPunct{\mcitedefaultmidpunct}
{\mcitedefaultendpunct}{\mcitedefaultseppunct}\relax
\EndOfBibitem
\bibitem[Rohlfing(2010)]{rohlfing_electronic_2010}
Rohlfing,~M. Electronic excitations from a perturbative {$\text{LDA}+GdW$}
  approach. \emph{Physical Review B} \textbf{2010}, \emph{82}, 205127\relax
\mciteBstWouldAddEndPuncttrue
\mciteSetBstMidEndSepPunct{\mcitedefaultmidpunct}
{\mcitedefaultendpunct}{\mcitedefaultseppunct}\relax
\EndOfBibitem
\bibitem[Winther and Thygesen(2017)Winther, and Thygesen]{winther_band_2017}
Winther,~K.~T.; Thygesen,~K.~S. Band structure engineering in van der Waals
  heterostructures via dielectric screening: the {G{$\Delta$}W} method.
  \emph{{2D} Materials} \textbf{2017}, \emph{4}, 025059\relax
\mciteBstWouldAddEndPuncttrue
\mciteSetBstMidEndSepPunct{\mcitedefaultmidpunct}
{\mcitedefaultendpunct}{\mcitedefaultseppunct}\relax
\EndOfBibitem
\bibitem[Ulstrup \latin{et~al.}(2016)Ulstrup, Čabo, Miwa, Riley, Grønborg,
  Johannsen, Cacho, Alexander, Chapman, Springate, Bianchi, Dendzik, Lauritsen,
  King, and Hofmann]{ulstrup_ultrafast_2016}
Ulstrup,~S.; Čabo,~A.~G.; Miwa,~J.~A.; Riley,~J.~M.; Grønborg,~S.~S.;
  Johannsen,~J.~C.; Cacho,~C.; Alexander,~O.; Chapman,~R.~T.; Springate,~E.;
  Bianchi,~M.; Dendzik,~M.; Lauritsen,~J.~V.; King,~P. D.~C.; Hofmann,~P.
  Ultrafast Band Structure Control of a Two-Dimensional Heterostructure.
  \emph{{ACS} Nano} \textbf{2016}, \emph{10}, 6315--6322\relax
\mciteBstWouldAddEndPuncttrue
\mciteSetBstMidEndSepPunct{\mcitedefaultmidpunct}
{\mcitedefaultendpunct}{\mcitedefaultseppunct}\relax
\EndOfBibitem
\bibitem[Gao and Yang(2017)Gao, and Yang]{gao_renormalization_2017}
Gao,~S.; Yang,~L. Renormalization of the quasiparticle band gap in doped
  two-dimensional materials from many-body calculations. \emph{Physical Review
  B} \textbf{2017}, \emph{96}, 155410\relax
\mciteBstWouldAddEndPuncttrue
\mciteSetBstMidEndSepPunct{\mcitedefaultmidpunct}
{\mcitedefaultendpunct}{\mcitedefaultseppunct}\relax
\EndOfBibitem
\bibitem[Liu \latin{et~al.}(2019)Liu, Ziffer, Hansen, Wang, and
  Zhu]{liu_direct_2019}
Liu,~F.; Ziffer,~M.~E.; Hansen,~K.~R.; Wang,~J.; Zhu,~X. Direct Determination
  of Band-Gap Renormalization in the Photoexcited Monolayer
  {${\mathrm{MoS}}_{2}$}. \emph{Physical Review Letters} \textbf{2019},
  \emph{122}, 246803\relax
\mciteBstWouldAddEndPuncttrue
\mciteSetBstMidEndSepPunct{\mcitedefaultmidpunct}
{\mcitedefaultendpunct}{\mcitedefaultseppunct}\relax
\EndOfBibitem
\bibitem[Cunningham \latin{et~al.}(2017)Cunningham, Hanbicki, {McCreary}, and
  Jonker]{cunningham_photoinduced_2017}
Cunningham,~P.~D.; Hanbicki,~A.~T.; {McCreary},~K.~M.; Jonker,~B.~T.
  Photoinduced Bandgap Renormalization and Exciton Binding Energy Reduction in
  {WS$_2$}. \emph{{ACS} Nano} \textbf{2017}, \emph{11}, 12601--12608\relax
\mciteBstWouldAddEndPuncttrue
\mciteSetBstMidEndSepPunct{\mcitedefaultmidpunct}
{\mcitedefaultendpunct}{\mcitedefaultseppunct}\relax
\EndOfBibitem
\bibitem[Fan \latin{et~al.}(2017)Fan, Zheng, Liu, Zhuang, Fan, Gong, Li, Wu,
  Jiang, Zhu, Zhang, Zhou, Hu, Wang, Duan, and Pan]{fan_nonlinear_2017}
Fan,~X.; Zheng,~W.; Liu,~H.; Zhuang,~X.; Fan,~P.; Gong,~Y.; Li,~H.; Wu,~X.;
  Jiang,~Y.; Zhu,~X.; Zhang,~Q.; Zhou,~H.; Hu,~W.; Wang,~X.; Duan,~X.; Pan,~A.
  Nonlinear photoluminescence in monolayer {WS} 2 : parabolic emission and
  excitation fluence-dependent recombination dynamics. \emph{Nanoscale}
  \textbf{2017}, \emph{9}, 7235--7241\relax
\mciteBstWouldAddEndPuncttrue
\mciteSetBstMidEndSepPunct{\mcitedefaultmidpunct}
{\mcitedefaultendpunct}{\mcitedefaultseppunct}\relax
\EndOfBibitem
\bibitem[Xie \latin{et~al.}(2019)Xie, Zhang, Li, Dong, Zhang, Wang, Liu,
  Kislyakov, Nunzi, Qi, Zhang, and Wang]{xie_layer-modulated_2019}
Xie,~Y.; Zhang,~S.; Li,~Y.; Dong,~N.; Zhang,~X.; Wang,~L.; Liu,~W.;
  Kislyakov,~I.~M.; Nunzi,~J.-M.; Qi,~H.; Zhang,~L.; Wang,~J. Layer-modulated
  two-photon absorption in {MoS$_2$:} probing the shift of the excitonic dark
  state and band-edge. \emph{Photonics Research} \textbf{2019}, \emph{7},
  762--770\relax
\mciteBstWouldAddEndPuncttrue
\mciteSetBstMidEndSepPunct{\mcitedefaultmidpunct}
{\mcitedefaultendpunct}{\mcitedefaultseppunct}\relax
\EndOfBibitem
\bibitem[Kang \latin{et~al.}(2017)Kang, Kim, Ryu, Jung, Kim, Moreschini,
  Jozwiak, Rotenberg, Bostwick, and Kim]{kang_universal_2017}
Kang,~M.; Kim,~B.; Ryu,~S.~H.; Jung,~S.~W.; Kim,~J.; Moreschini,~L.;
  Jozwiak,~C.; Rotenberg,~E.; Bostwick,~A.; Kim,~K.~S. Universal Mechanism of
  Band-Gap Engineering in Transition-Metal Dichalcogenides. \emph{Nano Letters}
  \textbf{2017}, \emph{17}, 1610--1615\relax
\mciteBstWouldAddEndPuncttrue
\mciteSetBstMidEndSepPunct{\mcitedefaultmidpunct}
{\mcitedefaultendpunct}{\mcitedefaultseppunct}\relax
\EndOfBibitem
\end{mcitethebibliography}

\end{document}


This pdf contains:
\begin{itemize}
\item[]\textbf{Figure S1:} Absorption spectra of \mose{},\ws{}, and \wse{} and corresponding absorption coefficients.
\item[]\textbf{Figure S2:} Density-dependent absorption spectra, band structure, and occupations of \mos{}.
\item[]\textbf{Supplementary Text:} Convergence analysis of the excited carrier density.
\item[]\textbf{Figure S3:} Convergence analysis of the density in \ws{} with different $\bk$-samplings.
\item[]\textbf{Table  S1:} Pump energies used in the density calculations.
\end{itemize}
\newpage
\renewcommand\thefigure{S\arabic{figure}}
\setcounter{figure}{0}
\renewcommand\thetable{S\arabic{table}}
%
\begin{figure*}[h!t]
\centering
\includegraphics[trim =0cm 8cm 0cm 8cm, clip, width=\textwidth]{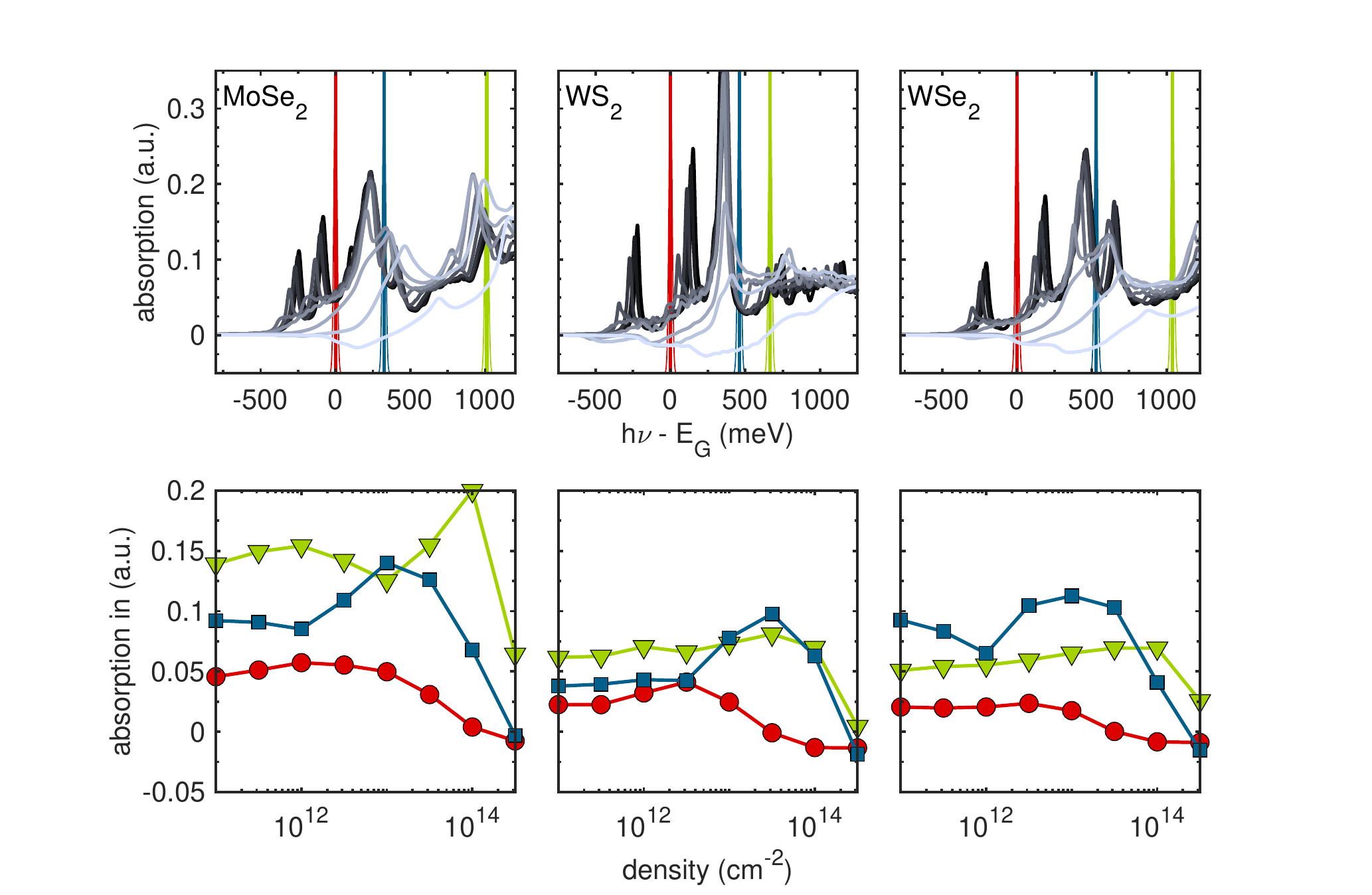}
\caption{Absorption spectra of \mose{},\ws{}, and \wse{} for a given excited carrier density in thermal equilibrium at 300K together with the energetic position and spectral width of the considered pump pulses (top row). The carrier density increases from zero (black line) to $3 \times 10^{14}$ cm$^{-2}$ (gray line). Intermediate carrier-density values correspond to the symbols in the bottom figures. Corresponding extracted absorption coefficients vs. carrier density at the three pump energies (bottom row).}
\label{fig:figA1}
\end{figure*}
\newpage
\begin{figure*}[h!t]
\centering
\includegraphics[trim =0cm 11cm 0cm 11cm, clip, width=\textwidth]{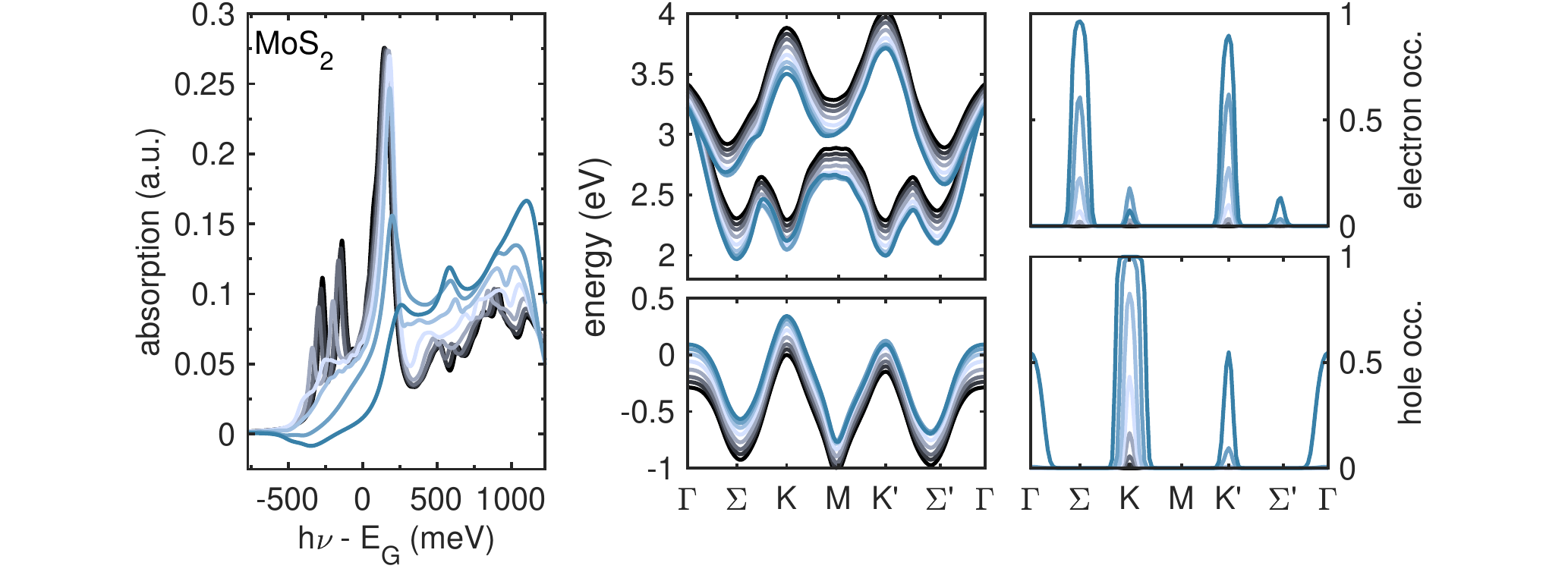}
\caption{Absorption spectrum, renormalized band structure, and photoexcited carrier occupation function across the Brillouin zone of \mos{} for a given excited carrier density in thermal equilibrium at 300K. The carrier density increases are zero (black line), $1.0 \times 10^{11}$ cm$^{-2}$, $3.6 \times 10^{11}$ cm$^{-2}$, $1.0 \times 10^{12}$ cm$^{-2}$, $3.6 \times 10^{12}$ cm$^{-2}$, $1.0 \times 10^{13}$ cm$^{-2}$, $3.6 \times 10^{13}$ cm$^{-2}$, $1.0 \times 10^{14}$ cm$^{-2}$, and $3.6 \times 10^{14}$ cm$^{-2}$ (dark blue line).
}
\label{fig:figA2}
\end{figure*}
\newpage
\subsection*{Convergence analysis of the excited carrier density}
\begin{figure*}[h!t]
\centering
\includegraphics[trim =0cm 10.5cm 0cm 11cm, clip, width=\textwidth]{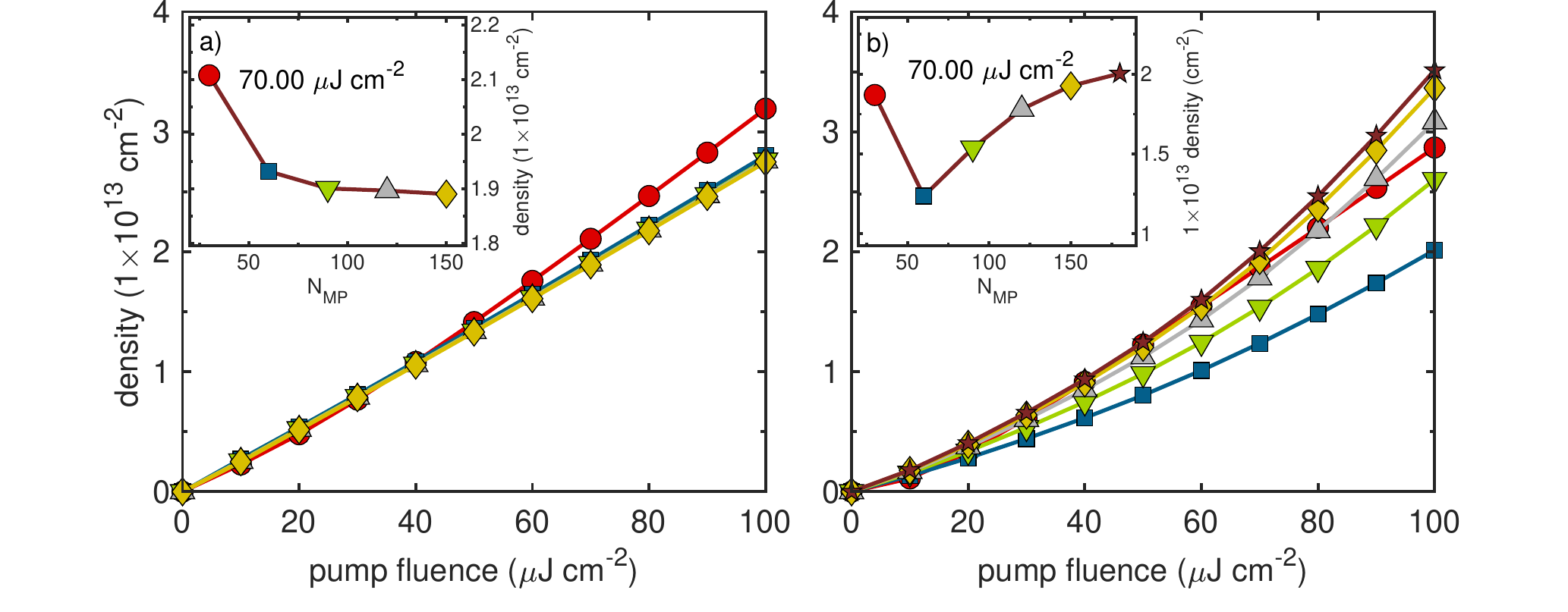}
\caption{Convergence analysis of the charge carrier density with different Gamma-centered Monkhorst-Pack $\bk$-point samplings using a $N_{MP} \times N_{MP} \times 1$ grid for \textbf{a)}: pumping at the TiSa energy and \textbf{b)}: pumping above the C-exciton in \ws{}. Results are compared for $N_{MP} = 30$ (red circles), $60$ (blue squares), $90$ (green triangles), $120$ (grey triangles), $150$ (yellow diamonds), and $180$ (brown stars). The insets demonstrate the convergence of the carrier density for a fixed pump fluence of 70 $\mu$J cm$^{-2}$.}
\label{fig:figA3}
\end{figure*}

%
The pump-induced carrier densities are calculated for \mos{}, \mose{}, \ws{}, and \wse{} using three different pump energies, as collected in Tab.~\ref{table:pump}, by numerically solving the semiconductor Bloch equations. The first Brillouin zone is sampled using a Gamma-centered Monkhorst-Pack grid. Sufficiently converged results are obtained for \mos{}, \mose{}, and \wse{} at all pump energies using a $(120\times 120\times 1)$ grid as exemplarily shown in Fig.~\ref{fig:figA3} (a). In \ws{}, strong excitation-induced modifications of the two-particle density-of-states in the vicinity of the C-exciton in combination with small intrinsic line broadening leads to slower convergence for pumping in this spectral region, see Fig.~\ref{fig:figA3} (b). For this specific case, a $(180\times 180\times 1)$ grid was used.
\begin{table}[h!t]
\begin{tabular}{c|c|c|c|c}
 & \mos{} & \mose{} & \ws{} & \wse{} \\
\hline
\hline
$E_G$ & 2292 & 2050 & 2396 & 2020 \\
Pump above C & 2624 & 2375 & 2856 & 2549 \\
TiSa laser & 3061 & 3061 & 3061 & 3061
\end{tabular}
\caption{Used pump energies in meV for the investigated materials. Here, $E_G$ is the quasi-particle band gap obtained from DFT+GW calculations for the respective freestanding monolayers, corrected by $G \Delta W$ shifts due to static screening from the hBN encapsulation layers.}
\label{table:pump}
\end{table}
%
%